\documentclass[11pt]{article}
\usepackage{amsfonts, amsmath, amssymb, esint, epsfig, graphicx, mathtools}
\usepackage[margin=1in]{geometry}
\usepackage{setspace}
\usepackage{hanging}
\usepackage{color}
\usepackage{verbatim, rotating}
\usepackage{framed}
\usepackage{enumerate}
\usepackage{tabularx}
\usepackage{booktabs} 
\usepackage[labelfont = bf]{caption} 
\usepackage[flushmargin, hang]{footmisc}  
\usepackage[hidelinks]{hyperref} 
\usepackage{dcolumn} 
\usepackage{pdflscape} 
\usepackage{titling} 
\usepackage{arydshln} 
\usepackage{float}
\usepackage{threeparttable} 
\usepackage[colorinlistoftodos]{todonotes}
\usepackage{caption,subcaption}
\usepackage{amsmath}
\usepackage{csquotes}

\usepackage[style=authoryear,backend=biber,natbib=true,isbn=false,uniquename=false]{biblatex}

\usepackage{multirow}
\usepackage{tcolorbox}

\bibliography{mybib}

\AtEveryBibitem{\clearfield{doi}}
\AtEveryBibitem{\clearfield{month}}
\AtEveryBibitem{\clearfield{urlmonth}}
\AtEveryBibitem{\clearfield{urlyear}}
\AtEveryBibitem{\clearlist{language}}
\renewbibmacro{in:}{%
  \ifentrytype{article}{}{\printtext{\bibstring{in}\intitlepunct}}}
\AtEveryBibitem{%
\ifentrytype{article}{\clearfield{url}}%
}
 
\newcommand{\revised}[1]{#1}
\newcommand{\revisedcolor}{}
\newcommand{\revisedcoloroff}{}

\usepackage{bbm}

\newtheorem{assumption}{Assumption}

\newcommand{\PreserveBackslash}[1]{\let\temp=\\#1\let\\=\temp}
\newcolumntype{C}[1]{>{\PreserveBackslash\centering}p{#1}}
\newcolumntype{R}[1]{>{\PreserveBackslash\raggedleft}p{#1}}
\newcolumntype{L}[1]{>{\PreserveBackslash\raggedright}p{#1}}

\makeatletter
\def\blfootnote{\xdef\@thefnmark{}\@footnotetext}
\makeatother

\usepackage{enumitem}
  {\begin{description}[leftmargin=*,before=]}
  {\end{description}}

\input{partialinput}

\usepackage{array}
\newcolumntype{H}{>{\setbox0=\hbox\bgroup}c<{\egroup}@{}}

\usepackage{authblk}

\begin{document}

\begin{titlepage}
\singlespace
\title{Optimal multi-action treatment allocation:\\ A two-phase field experiment to boost immigrant naturalization}

\author[1]{Achim Ahrens\thanks{\emph{Correspondence:} Achim Ahrens, Immigration Policy Lab, ETH Zurich. Leonhardshalde 21, CH--8092, +41 44 632 82 06. \emph{Emails:} \url{achim.ahrens@gess.ethz.ch} (Ahrens), \url{alessandra.stampi@gess.ethz.ch} (Stampi-Bombelli),
\url{dominik.hangartner@gess.ethz.ch} (Hangartner),
\url{selina.kurer@gess.ethz.ch} (Kurer). \emph{Acknowledgements:}
We thank our partners from the City of Zurich for their collaboration and support. We are grateful to Zhengyuan Zhou who has provided helpful feedback and to Teresa Freitas Monteiro for comments. We also thank seminar participants at ETH Zurich, Princeton University and Stanford University as well as participants at the PolMeth Europe conference. All remaining errors are our own. \emph{Funding:}
 This research was supported by the Stiftung Mercator Schweiz and by the \emph{nccr – on the move} program, which is funded by the Swiss National Science Foundation (grant no. 51NF40-182897).
\emph{Data:} The authors provide replication code through the Journal of Applied Econometrics Data Archive. The data is owned by the City of Zurich and the Canton of Zurich. It is not publicly available for confidentiality reasons. Data access requests should be directed to the City of Zurich's population office and the Canton of Zurich's municipal office.}}
\author[1]{Alessandra Stampi-Bombelli}
\author[1]{Selina Kurer}
\author[1]{Dominik Hangartner}
\affil[1]{Immigration Policy Lab, ETH Z\"urich}
\date{\today}
\maketitle

\begin{abstract}                    
\small
\revised{Research underscores the role of naturalization in enhancing immigrants' socio-economic integration, yet application rates remain low. We estimate a policy rule for a letter-based information campaign encouraging newly eligible immigrants in Zurich, Switzerland, to naturalize. The policy rule assigns one out of three treatment letters to each individual, based on their observed characteristics. We field the policy rule to one-half of 1,717 immigrants, while sending random treatment letters to the other half. Despite only moderate treatment effect heterogeneity, the policy tree yields a larger, albeit insignificant, increase in application rates compared to assigning the same letter to everyone.}


\medskip

\end{abstract}
\thispagestyle{empty}
\end{titlepage}



\section{Introduction}

Policymakers frequently need to select among alternative treatment options. While one of the stated aims of empirical research is to provide new insights to inform decision-making processes, the primary focus is usually on estimating averages of treatment effects rather than providing direct guidance on how to design assignment mechanisms for alternative treatments. In practice, the empirical researcher specifies a statistical model and estimates the efficacy of each treatment using an experimental or observational sample, while the decision maker assigns the treatment, interpreting the point estimates as if they were true. This approach, termed \emph{as-if} maximization by \citet{manski2021}, tends to yield one-size-fits-all rules assigning the same treatment to the wider population. Such one-size-fits-all policies seem inefficient given that treatment effects frequently exhibit relevant effect heterogeneity across observations and the increasing availability of administrative data providing rich individual characteristics. 

Policy learning provides a framework for directly estimating statistical decision rules, so-called policy rules, which prescribe treatments to individuals based on their observed characteristics (also known as profiling or targeting). While its origins date back to statistical decision theory \citep{wald1950statistical,savage1951}, the seminal work of \citet{manski2004} sparked a flourishing literature in econometrics which has developed methods for estimating statistical treatment rules, initially focusing on data drawn from randomized control trials \citep{manski2004,stoye2009,stoye2012,hirano2009}, but subsequently also covering observational data under unconfoundedness assumptions (\citealp{manski2007,athey2021a,zhou2022}; see \citealt{hirano2020} for a review). While applied research using policy learning is still relatively scarce, previous work has revealed the potential for data-driven treatment allocation across a variety of domains, including active labor market programs \citep[e.g.][]{lechner2007,frolich2008a}, vaccines accounting for spill-over effects \citep{kitagawa2023}, deforestation-reducing policies \citep{assuncao2022}, anti-malaria subsidies under budget constraints \citep{bhattacharya2012}, energy use information campaigns \citep{ida2022,gerarden2022} and maximizing fundraising \citep{cagala2021}.  

In this pre-registered study, we co-design and evaluate an individualized treatment allocation program with the goal of facilitating the naturalization of eligible immigrants in the City of Zurich, Switzerland. An expanding body of literature is utilizing close naturalization referendums or temporal discontinuities created by policy reform to enable credible comparisons between naturalized and non-naturalized immigrants to demonstrate that acquiring host-country citizenship offers long-term integration benefits for immigrants, and, indirectly, to host societies. These benefits span various integration dimensions, including employment and earnings \citep{gathmann2018access, hainmueller2019effect}, political efficacy, knowledge, and participation \citep{hainmueller2015naturalization}, as well as social incorporation and cooperation \citep{keller2015citizenship, hainmueller2017catalyst, felfe2021more}. Yet, despite these benefits, naturalization rates remain low in many countries, with a median annual naturalization rate (number of naturalized immigrants divided by number of immigrants) of 1.9\% in Europe and 3.1\% in the U.S. \citep{ward2019large}. Against this background, policymakers at the national level in Estonia, Ireland, Latvia, North Macedonia, Spain, and at the local level in Germany, Italy, Switzerland, and the United States, have deployed information campaigns to boost citizenship applications  and fully reap the integration benefits of naturalization \citep{huddleston2013naturalisation}.

Informed by existing research \citep[e.g.,][]{bloemraad2002north, baubock_acquisition_2006, national2016integration} and the insights of integration and naturalization bureaucrats of the City of Zurich, this study considers interventions that address three specific hurdles blocking eligible immigrants' path to citizenship. These hurdles include: (i) the perceived complexity of the application process, (ii) knowledge gaps about the requirements for naturalization, and (iii) the feeling of not being welcome to naturalize. To address the first two hurdles, we co-designed specific information letters with the City of Zurich. For the third hurdle, a letter sent by the  Mayor of City of Zurich encouraged immigrants to apply. In line with recent recommendations by \citet{haaland2023}, we opted for three separate treatment letters with accompanying flyers to ensure that each letter is short and easy to understand. Addressing all hurdles in a combined treatment letter with several flyers is likely counterproductive due to the limited time and attention that recipients devote when reading the letters.\footnote{A large literature in behavioral economics stresses that information processing is costly and provides evidence that individuals often fail to translate all available information into optimal decisions (for recent reviews, see \citealp{handel2018,gabaix2019,mackowiak2023}).} 

Since it is unknown which treatment letter is optimal for maximizing the individual application probabilities, and given that the optimal treatment choice may differ among individuals, we derive a multi-action policy rule. This policy rule is structured as a decision tree, which is referred to as a `policy tree.' Policy trees are introduced by \citet{athey2021a} for binary and by \citet{zhou2022} for multi-valued treatments. In our context, the policy tree selects one treatment from a set of three treatment options for each eligible immigrant based on their individual characteristics including residency, nationality and age. The treatment options are incorporated into three different letters with enclosed flyers sent out by the City of Zurich. Thus, by applying policy learning, we allow the optimal content and framing of the information provision to vary with observed immigrant characteristics. The policy rule is chosen to maximize the application rate for naturalization, the first step in the process of acquiring Swiss citizenship. 

Policy trees possess several strengths that make them a particularly promising method for immigrant naturalization and other sensitive policy contexts. First, policy trees allow policymakers and researchers to select those variables that can be used to tailor treatment assignment and, more importantly, exclude those that should not be used (e.g., protected characteristics such as religion)---and quantify the costs of exclusion in terms of foregone treatment efficacy. Second, policy trees make transparent which variables, and which variable values, guide treatment assignment. This is in contrast to black-box \emph{plug-in} rules, providing no insights into what drives treatment allocation. Related to the second strength is the third: policy trees are easy to visualize and easy to explain to users of the research---e.g., policymakers, case officers, and subjects receiving treatment assignment---even if they lack training in statistics. Together, transparency and interpretability are important steps towards satisfying requirements for explainable Artificial Intelligence (AI), e.g., as outlined in recent proposals for the regulation of AI by the \citet[][]{european2021laying} and \citet{thewhitehouse2022}. Finally, from a practical perspective, the so-called offline approach of policy trees, which learns policies from a single data batch, is often easier to implement in a public policy context than adaptive approaches training policy rules dynamically over time (e.g., \cite{caria2020adaptive}).

\revised{After introducing the methodology of policy learning, we} illustrate the practical feasibility of the targeted assignment rule and evaluate its benefits using a tailored, two-phase randomized controlled trial. In the first phase \revised{of our field experiment}, we randomly allocate 60\% of our sample of 5,145 citizenship-eligible immigrants to receive one of three letters addressing specific naturalization hurdles. Based on first-wave application outcomes and leveraging observed treatment effect heterogeneity, we estimate the optimal multi-action policy rule using the estimation framework of \citet{zhou2022}. In the second phase, we field the fitted policy rule on one-half of the remaining sample while sending random treatment letters to the other half. Adopting terminology from reinforcement learning, we refer to these two phases as the exploration phase (aimed at gathering knowledge about treatment efficacy) and the exploitation phase (aimed at implementing the reward-maximizing strategy), respectively. We evaluate the performance of the derived policy rule against random treatment allocation, one-size-fits-all policy rules assigning the same treatment to everyone, and a model-free \emph{plug-in} rule assigning the treatment with the largest estimated treatment effect. We find that policy trees can capture the vast majority of treatment effect heterogeneity of the more flexible but less transparent and non-interpretable \emph{plug-in} rule. Despite only moderate levels of heterogeneity, the policy tree yields a larger, albeit insignificant, increase in take-up than each individual treatment. 

Our study relates to three fields of empirical research. First, sparked by methodological advances, especially the advent of causal forests \citep[due to][]{wager2018}, there is a burgeoning literature estimating heterogeneous treatment effects using machine learning \citep[e.g.,][]{davis2017using,knittel2021,knaus2022}.\footnote{Other methods for estimating conditional average treatment effects using machine learning include \citet{Chern2018_genericml,kunzel2019a}. For an overview, see \citet{knaus2021,jacob2021a}.} While studies in this literature emphasize the potential of estimating heterogeneous effects for improved targeting, they usually do not explicitly derive interpretable targeting rules. Second, we build on the expanding literature applying statistical decision rules. The vast majority of applied studies, including those discussed above (i.e., \citealp{lechner2007,frolich2008a,bhattacharya2012,assuncao2022,kitagawa2023}), only provide backtest results about the ex-post performance of policy targeting rules. \citet{ida2022} propose a policy-learning framework that allows participants to self-select their treatment and apply their method to a residential energy rebate program. Closest to our study are \citet{gerarden2022} and \citet{cagala2021}. \citet{gerarden2022} follow the methodology of \citet{kitagawa2018} to estimate policy rules for a behavioral intervention targeted at reducing household electricity usage, but do not implement the derived policy rules. Similar to us, \citet{cagala2021} consider policy trees in an application to maximizing fundraising and gauge the performance of the estimated policy tree on out-of-sample data. We add to this literature by fielding the estimated optimal policy rule in the second phase of our experiment, which allows us to directly evaluate the performance against other policy rules. Furthermore, both \citet{cagala2021} and \citet{gerarden2022}  focus on the choice between two treatment options, whereas we are concerned with the more challenging problem of multi-action policy learning. Third, we contribute to the larger literature on informational interventions aimed at increasing take-up of government services and subsidies among eligible people \citep[e.g.,][]{bhargava2015psychological,finkelstein2019,hotard2019low, goldin2022tax}. \revised{Beyond contributing to these three strands, this article aims to make policy learning accessible to a wider audience by offering an introduction relevant both for randomized field experiments and applications relying on observational data.} 

This article proceeds as follows. \revised{Section~\ref{sec:policy_learning} provides an introduction to} policy learning. Section~\ref{sec:design} turns to our application. We contextualize our application, describe the data, the treatments and the study design in Sections~\ref{sec:background}-\ref{sec:study_design}. We summarize the results of the exploration and exploitation phase in Sections~\ref{sec:wave_1_results} and \ref{sec:wave_2_results}. Section~\ref{sec:conclusion} concludes.

\section{Multi-action policy learning} \label{sec:policy_learning}

In this section, we provide a brief review of (multi-action) policy learning, with a special focus on the policy learning framework of \citet{zhou2022}. While we rely on a randomized experimental design to learn the optimal policy rule in our application, we also discuss the setting where one has to rely on unconfoundedness assumptions, thereby illustrating the generality of the methodological framework.

The aim of policy learning is to formulate a policy rule $\pi(X)$ designed to maximize the expected value of $Y$, the outcome of interest. A policy rule assigns a treatment $a$ from the choice set of treatment options $\mathcal{A}=\{1,2,\dots,D\}$ to each individual based on their observed covariates $X$. Note that $\mathcal{A}$ may include the no-treatment option. Formally, $\pi(X)$ is a function mapping individual characteristics to one of the treatment options in $\mathcal{A}$. For example, a policy rule might assign treatment~1 to every person below age 30, treatment~2 to individuals aged 30-40, and treatment~3 to individuals older than 40. 

\subsection{\revised{Estimating optimal policies}}
Before we turn to the estimation of optimal policies, it is instructive to consider a candidate policy rule \revised{$\pi^\prime(X)$} and assess its effectiveness. We assume that we have access to \revised{the sample $\{Y_{i},A_{i},X_{i}\}$ for $i=1,\ldots,n$, which is drawn from the joint population distribution $P$. The sample data include} the treatment received, $A_{i}$, the realized outcome, $Y_{i}$, as well as observed individual $i$'s characteristics $X_i$. In our application, the data stems from the exploration phase of the randomized controlled trial, but the general approach also extends to observational data.

As typical in the causal effects literature, we assume the existence of the potential outcomes $\{Y_{i}(1),Y_{i}(2),\ldots,Y_{i}(D)\}$, which are the outcomes if individual $i$ had received treatments 1, 2, \ldots, $D$ \citep{rubin1974,imbens2015causal}. This allows us to define the expected reward of  \revised{$\pi^\prime(X)$}, which is the expected value of the potential outcomes if the policy rule had been followed, i.e., $Q(\pi^\prime\revised{(X_i)})=E[Y_{i}(\pi^\prime(X_i))]$ \revised{where $E[\cdot]$ denotes the expectation with respect to the population $P$.} \revised{In non-experimental settings, the fundamental challenge of estimating the reward of a candidate policy \revised{$\pi^\prime(X)$} is that we only observe $Y_i=Y_i(A_i)$ and that individuals} might self-select into treatment options that optimize their expected pay-off. 

The offline policy learning literature commonly imposes the following set of assumptions \citep{kitagawa2018,zhou2022}:

\begin{assumption}
\begin{enumerate}[noitemsep,label=(\alph*)]
    \item Unconfoundedness: $Y_{i}(1),\ldots,Y_{i}(D)\} \perp A_i\vert X_i.$
    \item Overlap: There exists some $\eta>0$ such that $e_a(X_i) \geq \eta$ for any $a\in\mathcal{A}$ and $X$, where  $e_a(X_i) \equiv P(A_i=a\vert X_i)$ are the propensity scores for treatment $a$. 
    \item Boundedness: The potential outcomes are contained on a finite interval in $\mathbb{R}^D$. 
\end{enumerate}
\end{assumption}

Unconfoundedness in (a) states that we observe all necessary covariates allowing us to account for selection biases. The condition is naturally satisfied by randomized treatment assignments. 
The overlap assumption in (b) requires that for any observed individual characteristic $X_i$, the probability $e_a(X_i)$ of taking each action $a$ is greater than zero. The boundedness assumption in (c) serves the purpose of simplifying mathematical proofs but can be replaced by weaker assumptions. 

Under the stated assumptions, we can evaluate the reward of a candidate policy $\pi^\prime$ by averaging over observations that happen to align with the candidate policy rule, i.e., 
\begin{equation} \widehat{Q}_{IPW}(\pi^\prime\revised{(X_i)})=\frac{1}{n} \sum_{i=1}^n \frac{{\mathbbm{1}}{\left\{A_i=\pi^\prime\left(X_i\right)\right\}}Y_i}{e_{A_i}(X_i)} . \label{eq:qipw}\end{equation}
where we \revised{take the weighted average across observations that align with the candidate policy rule.} We inversely weight by the propensity score $e_a(X_i)$ to account for selection bias \citep{swaminathan2015,kitagawa2018}. 

Suppose that the policymaker suggests a number of policy rules, e.g., $\Pi^\prime=\{\pi^{\prime},\pi^{\prime\prime},\pi^{\prime\prime\prime}\}$ where $\Pi^\prime$ is the set of candidate policies. The optimal policy is the policy that maximizes the expected reward; formally, ${\pi}^\star\revised{(X_i)}=\arg\max_{\pi\in\Pi^\prime} Q(\pi\revised{(X_i)})$. Accordingly, we can leverage our sample to estimate the optimal policy rule as 
\begin{equation} \revisedcolor
\hat\pi(X_i)=\arg\max_{\pi\in\Pi^\prime} \hat{Q}_{IPW}(\pi\revised{(X_i)}). \label{eq:policy_est}
\end{equation}
The performance of a policy learner $\hat\pi\revised{(X_i)}$, which estimates $\pi^\star\revised{(X_i)}$ from the data, is measured by its regret, $R(\hat\pi\revised{(X_i)})=Q(\pi^\star\revised{(X_i)})-Q(\hat\pi\revised{(X_i)})$. Thus, regret measures the difference between the reward of the unobserved optimal policy and the value of the estimated policy.

\revised{Note that we leverage the same covariates $X_i$ to adjust for selection effects in~\eqref{eq:qipw} and to form policy rules in~ \eqref{eq:policy_est}. There may, however, be good reasons to use distinct covariate sets in each step. For example, legal or ethical concerns might mandate the exclusion of protected characteristics from the policy rule (e.g.,\ gender, nationality, religion). Yet, the inclusion of these characteristics in the propensity score estimation could be necessary if prior evidence suggests their potential influence on treatment allocation. In randomized experiments, a consistent estimation of the reward does not require covariate adjustment but can enhance statistical precision.}

\subsection{\revised{Cross-fitting and double-robust estimation}}

If the propensity scores $e_a(X)$ are known, the regret converges to zero at $\sqrt{n}$-rate \citep{swaminathan2015,kitagawa2018}. If the exact assignment mechanism is not known, \revised{which is typically the case in non-experimental settings}, we have to estimate $e_a(X)$ from the data. One approach is to estimate $e_a(X)$ using the full sample and plug the estimates into \eqref{eq:qipw}. \revised{However, this approach generally yields sub-optimal convergence rates unless we impose strong convergence rate requirements on the first-step estimator \citep{kitagawa2018}. The sub-optimal performance can be attributed to the own-observation bias, which arises if the first-step estimation error from the propensity score estimation is correlated with the error associated with estimating the reward.} 
To allow for \revised{a general class of} data-adaptive nonparametric estimators, including popular supervised machine learners such as random forests, which are more robust towards unknown data structures, \citet{zhou2022} combine two strategies for policy learning: cross-fitting and double-robust estimation using augmented inverse-propensity weighting (AIPW). We discuss each strategy in turn.

\revised{To illustrate how cross-fitting addresses the own-observation bias, consider the simple case where we randomly split the data into two sub-samples, referred to as auxiliary and main samples. In the first step, we leverage the auxiliary sample for the estimation of conditional expectation functions (e.g., the propensity scores). The second step uses out-of-sample predicted values from the first step on the main sample to estimate the reward. This sample-splitting approach resolves the own-observation bias since the second step is, after conditioning on the auxiliary sample, independent from the first-step estimation error.}

\revised{Cross-fitting extends this sample-splitting approach by flipping the auxiliary and main samples, thus effectively using the full sample for both the first and second-step estimation. Cross-fitting also allows the splitting of the data into more than two partitions.\footnote{The causal machine learning literature frequently relies on sample splitting approaches, such as cross-fitting; see for example \citet{chernozhukov2018b} for the estimation of average treatment effects and \citet{wager2018} for the estimation of CATE using causal forests.} Specifically,} to implement cross-fitting, we randomly split the sample into $K$ folds of approximately equal size. We use $\hat{e}_{a}^{-k(i)}(X_i)$ to denote the {\it cross-fitted} propensity score of observation $i$ for treatment $a$. The cross-fitted predicted value is calculated as the out-of-sample predicted value from fitting an estimator on all folds but fold $k(i)$, which is the fold that observation $i$ falls into. Similarly, we introduce $\hat{\mu}_{a}^{-k(i)}(X_i)$ which is the cross-fitted predicted value of the outcome under treatment $a$ using predictors $X_i$, i.e., it is a cross-fitted estimate of \revised{$\mu_a(X_i)\equiv E[Y_i(a)\vert X_i]$}. 

\revised{Double robust estimation of the reward allows for non-random treatment assignment under uncounfoundedness. The estimator adjusts for biases arising from selective treatment allocation by combining the reweighting approach of inverse-propensity weighting \revised{as used in \eqref{eq:qipw}} with outcome adjustment (as used in regression-based adjustment). The advantage over the IPW estimator is that the resulting double-robustness property guarantees consistency if either the propensity scores $e_a(X_i)$ or the conditional expectation of outcome given covariates, i.e., $\mu_a(X_i)$, are correctly specified.} Using the cross-fitted estimates $\hat{e}_{a}^{-k(i)}(X_i)$ and $\hat{\mu}_{a}^{-k(i)}(X_i)$, we can define the cross-fitted AIPW (CAIPW) estimator of the reward as\footnote{The function $\mathbbm{1}\{\cdot\}$ denotes the indicator function.}
\begin{equation}
\widehat{Q}_{CAIPW}(\pi\revised{(X_i)})=\frac{1}{n} \sum_{i=1}^n  
\left( \frac{Y_i-\hat{\mu}_{A_i}^{-k(i)}\left(X_i\right)}{\hat{e}_{A_i}^{-k(i)}\left(X_i\right)} \mathbbm{1}\{ A_i = \pi(X_i)   \}+ 
\hat{\mu}_{\pi(X_i)}^{-k(i)}\left(X_i\right)  \right).
\label{eq:caipwl}
\end{equation}
The first term in~\eqref{eq:caipwl} adjusts the observed outcome by subtracting the conditional expectation of the outcome under the observed treatment and by inversely weighting with the propensity scores \revised{if the observed treatment, $A_i$, is equal to the treatment recommended by the policy, $\pi(X_i)$. The} second term adds the conditional expectation of the outcome under the treatment assigned by the policy. \revised{Using the double-robust estimator of the reward, we can estimate the optimal policy by evaluating $\widehat{Q}_{CAIPW}(\pi\revised{(X_i)})$ for all candidate policies in $\Pi^\prime$, i.e., we calculate ${\hat\pi\revised{(X_i)}}_{CAIPW}=\arg\max_{\pi\in\Pi^\prime} \widehat{Q}_{CAIPW}(\pi\revised{(X_i)})$.}

\subsection{Policy class}
So far, we have assumed a predefined set of candidate policies. In many applications, however, we wish to learn policies flexibly from the data instead of relying on a pre-defined set of policy rules. A fully flexible approach could assign each individual to the treatment for which the estimated treatment effect is the largest. This {\it plug-in policy rule} requires no functional form restrictions but may be inappropriate when stakeholders wish to learn about the drivers of treatment efficacy and have hesitations to rely on a black-box treatment assignment mechanism.\footnote{For formal results on plug-in rules, see \citet{hirano2009,bhattacharya2012}.} 

\begin{figure}
    \centering\footnotesize
    \begin{subfigure}[t]{.4\linewidth}
        \includegraphics[width=\linewidth,trim={1.1cm 1.1cm 1cm 1cm},clip]{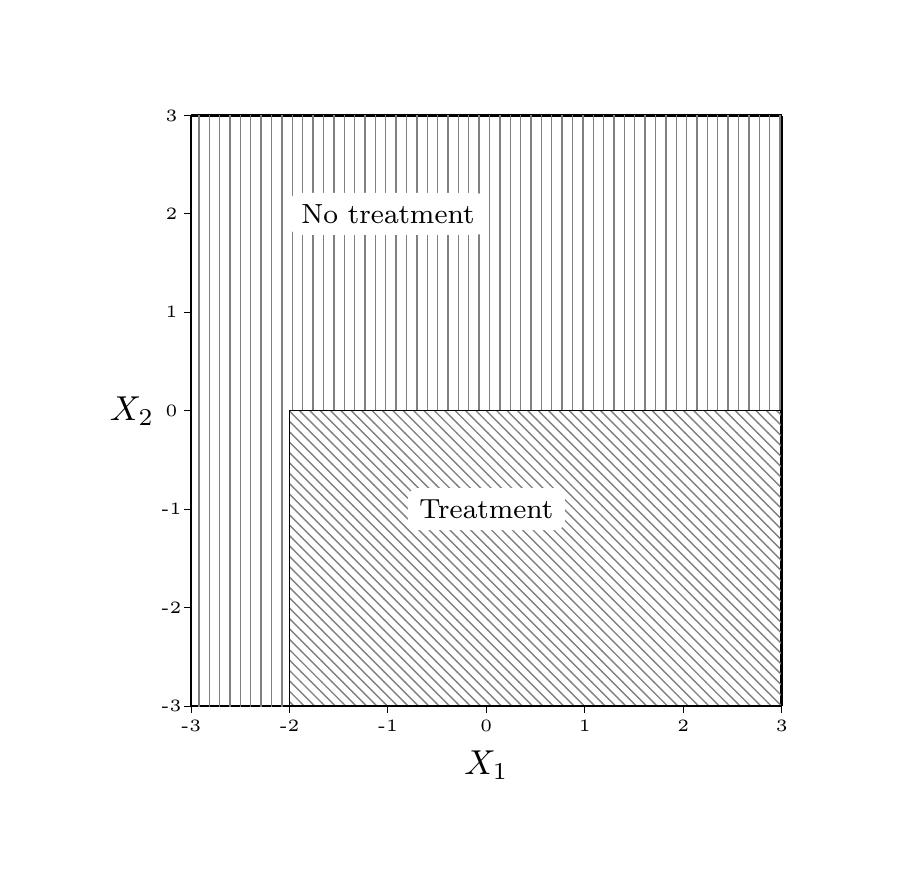}
        \caption{\revisedcolor Quadrant policy rule}
    \end{subfigure}\hspace{.05\linewidth}%
    \begin{subfigure}[t]{.4\linewidth}
        \includegraphics[width=\linewidth,trim={1.1cm 1.1cm 1cm 1cm},clip]{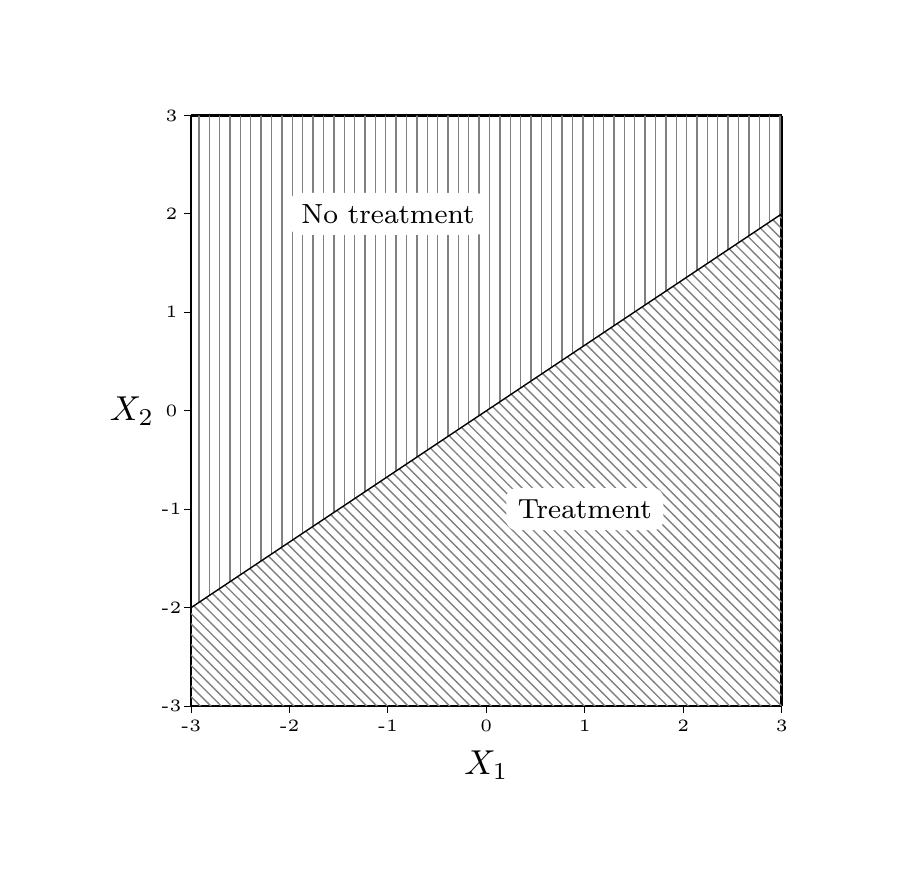}
        \caption{\revisedcolor Linear policy rule}
    \end{subfigure}\par\bigskip %
    \begin{subfigure}[t]{.4\linewidth}
        \includegraphics[width=\linewidth,trim={0cm -3cm 0cm 0cm},clip]{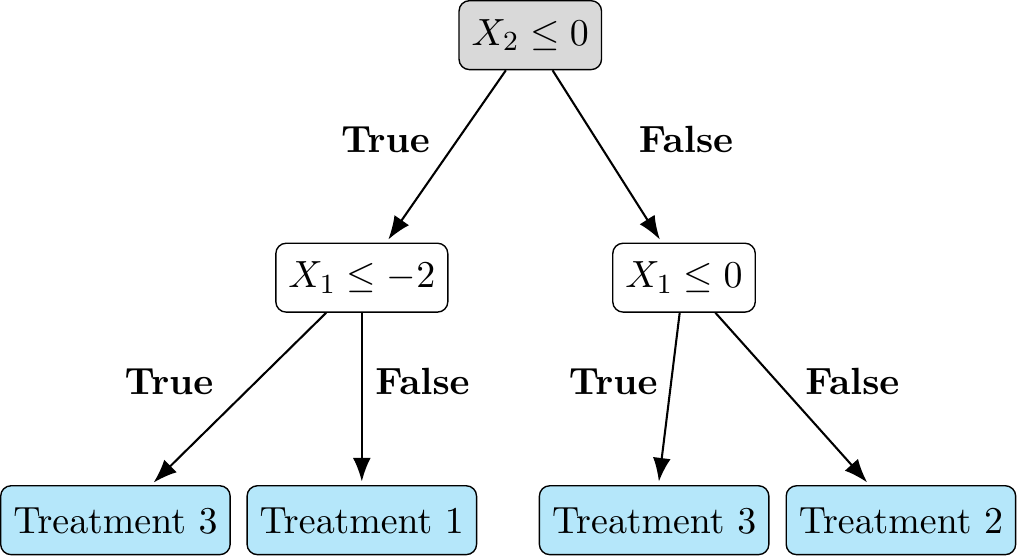}
        \caption{\revisedcolor Policy tree}
    \end{subfigure}\hspace{.05\linewidth}%
    \begin{subfigure}[t]{.4\linewidth}
        \includegraphics[width=\linewidth,trim={1.1cm 1.1cm 1cm 1cm},clip]{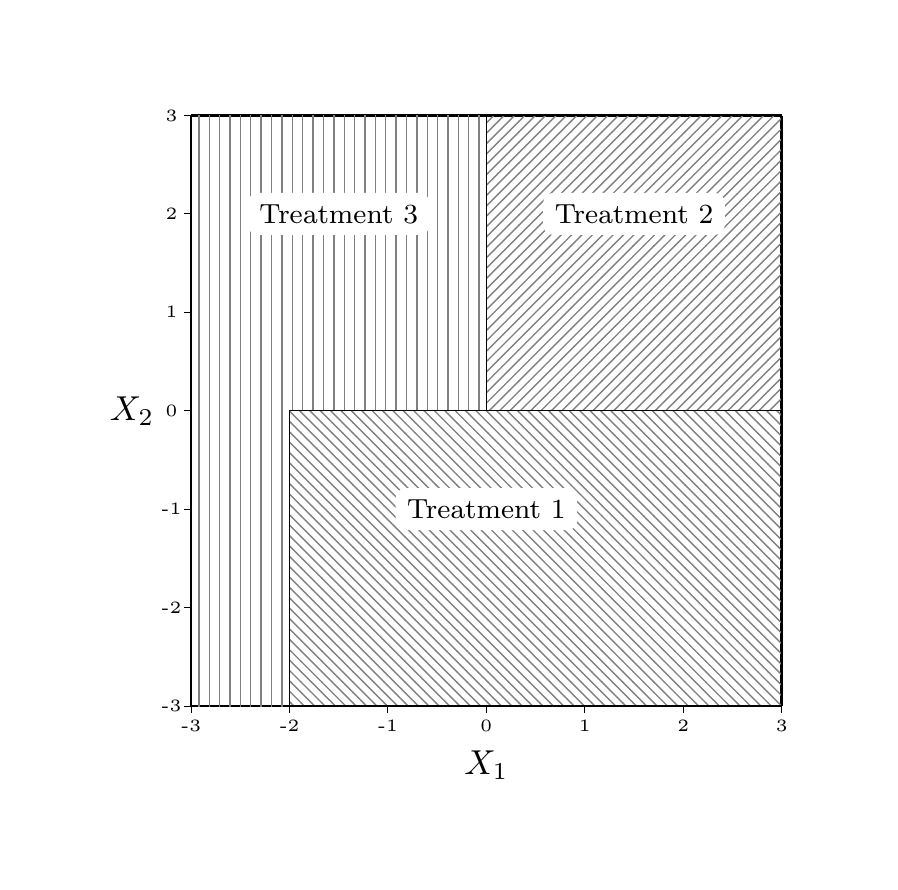}
        \caption{\revisedcolor Policy tree in two-dimensional space}
    \end{subfigure}\par\bigskip
    \parbox{\linewidth}{\revisedcolor \emph{Note:} The figure shows three examples of policy rules that assign treatments based on two covariates $X_1$ and $X_2$. Figures~(a) and (b) illustrate examples of the quadrant and linear policy rule considered in \citet{kitagawa2018} for binary treatment choices. Figure~(c) and (d) provide two alternative illustrations of the same policy tree with three treatment options.}
    \caption{\revisedcolor Illustrative examples of policy rules with different functional forms}
    \label{fig:policies_examples}
\end{figure}

Policy learning allows estimating interpretable treatment rules from the data. To this end, we must choose a suitable policy class from which we estimate the optimal policy. In an application to active labor market programs with a binary treatment and two covariates, \citet{kitagawa2018} discuss three policy classes defined by the three functional form restrictions:
\begin{alignat*}{2}
 \textnormal{Quadrant policy rule:}&&\qquad    \pi(X_i) &{=} \mathbbm{1}\{s_1(X_{1i}-\beta_1)\geq 0\}\mathbbm{1}\{s_2(X_{2i}-\beta_2)\geq 0\} \\
 \textnormal{Linear policy rule:} &&    \pi(X_i) &{=} \mathbbm{1}\{\beta_0+\beta_1X_{1i}+ \beta_2X_{2i}\geq 0\} \\ 
  \textnormal{Cubic policy rule:} &&   \pi(X_i) &{=} \mathbbm{1}\{\beta_0+\beta_1X_{1i}+ \beta_2X_{2i} + \beta_3X_{2i}^2+ \beta_4X_{2i}^3\geq 0\}  
\end{alignat*}
where $s_1,s_2\in \{-1,+1\}$ and $\beta_j\in \mathbbm{R}$. The first rule defines a quadrant in the two-dimensional space spanned by the covariates $X_1$ and $X_2$, and assigns the treatment to individuals for which $X_{1i}$ and $X_{2i}$ lie in that quadrant. The second rule defines a linear decision boundary. \revised{Figure~\ref{fig:policies_examples}(a) and (b) illustrate examples of the quadrant and linear policy rule. T}he third rule allows for non-linear decision boundaries by including quadratic and cubic terms. Compared to the plug-in rule, these rules exhibit a higher degree of interpretability \revised{and, as shown in Figure~\ref{fig:policies_examples}, they can} be easily visualized in a two-dimensional plot.

In a multi-action setting, \citet{zhou2022} focus on the class of policy rules that take the form of (shallow) decision trees. Trees are widely employed as predictive tools that construct predictions by splitting the feature space into non-overlapping regions. In the prediction context, classification and regression trees yield the same prediction for observations falling into the same region. In the policy context, observations falling into the same region are assigned the same treatment action. \revised{Figure~\ref{fig:policies_examples}(c) shows an example of a policy tree which assigns treatment options 1, 2 or 3 based on the values of $X_1$ and $X_2$. Figure~\ref{fig:policies_examples}(d) visualizes the same tree in a two-dimensional space, highlighting that policy} trees are conceptually similar to the quadrant rule but can be generalized to multiple treatments. \citet{zhou2022} describe how the optimization of policy trees can be regarded as a mixed integer program and \citet{policytree} implement a less costly but approximate optimization algorithm, referred to as {\it hybrid} tree search.

To conclude, the policy learning framework of \citet{zhou2022} has two attractive properties, \revised{which are particularly relevant in the context of the} the naturalization campaign \revised{discussed} in the next section: First, since trees can be easily visualized, they are transparent and simple to interpret, even without statistical training, making them attractive in a public policy context where users of the research vary in statistical literacy and often view black-box methods with scepticism. Second, the estimation of policy trees is computationally feasible with readily available software \citep{grf,policytree}. 


\section{Personalizing naturalization campaigns}\label{sec:design}

In this section, we apply the multi-action policy tree to an information campaign encouraging eligible immigrants residing in the City of Zurich to apply for Swiss citizenship. We introduce the policy context in Section~\ref{sec:background}, and discuss data and treatments in Section~\ref{sec:data} and \ref{sec:treatments}. Section~\ref{sec:study_design} summarizes the study design and estimation methodology. Results are presented in Section~\ref{sec:wave_1_results} and~\ref{sec:wave_2_results}.

\subsection{Background: Immigrant Integration and Citizenship} \label{sec:background}
The integration of immigrants into the host-country fabric and economy is a central policy issue in many countries across the globe. One promising policy to foster integration is naturalization, i.e., the process of awarding host-country citizenship to immigrants \citep{goodman2014immigration, dancygier_immigration_2010}. Observational studies relying on difference-in-difference models and regression discontinuity designs comparing similar naturalized and non-naturalized immigrants show that acquiring host-country citizenship can positively impact the integration of immigrants by increasing their earnings and labor market attachment \citep{oecd_naturalisation:_2011, mazzolari2009dual,gathmann2018access,hainmueller2019effect, vink2021long, govind2021naturalization, gathmann2023citizenship}, fostering political efficacy and knowledge \citep{hainmueller2015naturalization}, spurring cultural assimilation and cooperation \citep{keller2015citizenship, felfe2021more}, and reducing feelings of isolation and discrimination  \citep{hainmueller2017catalyst}.\footnote{With the exception of \citet{mazzolari2009dual}, who studies immigrants from Latin American countries in the U.S, the studies referenced above focus on France, Germany and Switzerland. This might limit external validity since we expect the benefits of naturalization to be context-dependent and generally decline with lower naturalization hurdles. While testing this hypothesis requires more comparative research, \citet{vernby2019can} provide initial evidence from a correspondence test varying citizenship in fictitious applications in Sweden that is consistent with this conjecture.} This process can also benefit the host society by increasing immigrants' contributions to economic growth, lowering their dependency on welfare, and, by extension, reducing societal tensions and strengthening social cohesion \citep[for reviews, see][]{national2016integration, pastor2012economic}.

Despite these potential benefits, naturalization rates remain low in many countries \citep{blizzard2019naturalization}. What explains this mismatch between the benefits of host-country citizenship and the low demand for naturalization? Previous evidence from surveys and qualitative studies suggest that uncertainty about the eligibility criteria such as residency and language requirements can prevent immigrants from applying \citep{baubock_acquisition_2006, gonzalez2013path}. Other studies highlight that---particularly in hostile immigration environments---a lack of encouragement by politicians, public administration, or the general public might deter immigrants \citep{baubock_acquisition_2006, bloemraad2002north, bloemraad_citizenship_2008}. Furthermore, in earlier research using a tailored survey, we find evidence for informational deficits and the feeling that an application is not welcome by the host society \citep{hangartner2023citizenship}. Lastly, in countries that unlike Switzerland do not allow for dual citizenship, immigrants might not be willing to give up the passport from their origin country to obtain host-country citizenship.\footnote{Whether a person is allowed to retain the previous citizenship when naturalizing in another country generally depends on the regulations of both the origin and host country. Switzerland has guaranteed the right to hold dual (or more) citizenship without restrictions since 1992. Hence, eligible immigrants seeking Swiss citizenship are subject only to restrictions of their origin countries. Among the ten largest countries by nationality in our sample (which jointly amount to 72\% of origin countries), only Austria and Spain generally do not allow for dual citizenship.}

To boost naturalization rates, countries, states, and municipalities across Europe and the U.S. have begun to turn to information campaigns to overcome hurdles to citizenship acquisition for eligible immigrants. While the content and scope of these naturalization campaigns vary, they often combine information provision about the naturalization process and requirements with an encouragement to apply for citizenship. Yet, despite the growing popularity of these campaigns across Europe and the U.S., there exists little experimental research to evaluate its effectiveness. An important exception is \citet{hotard2019low}, who show that a low-cost nudge informing low-income immigrants about their eligibility for a fee waiver increased the rate of citizenship applications by 8.6 percentage points (from 24.5\% in the control group to 33.1\%).  Most similar to our study is \citet{hangartner2023citizenship}, who evaluated previous versions of the naturalization campaign of the City of Zurich and showed that a similarly low-cost letter (about CHF 1.20 per person, see below) combining information and encouragement increased naturalization rates by about 2.5 percentage points (from 6.0\% in the control group to 8.5\%).

Past naturalization campaigns, including the one by the City of Zurich mentioned above, have typically relied on a one-size-fits-all approach---despite the substantial diversity of the immigrant population in terms of, e.g., country of origin, language skills, and age. There are good reasons to suspect treatment effect heterogeneity along various dimensions: Immigrants' willingness to naturalize and their susceptibility to certain information letters might depend on their current nationality due to the specific dual citizenship regulations, the relative benefits in terms of visa requirements \emph{vis-\'a-vis} third countries, the attachment to the home country, and the attitudes of native citizens towards specific immigrants groups. For example, immigrants who feel discriminated against might be more likely to be persuaded by a letter welcoming them to set roots and apply for citizenship in their host country. Furthermore, language requirements might be less of a concern for immigrants who speak the same language (such as Austrians and Germans in Switzerland). Thus, tailoring such campaigns to the specific needs of diverse immigrants promises to deliver both a deeper understanding of the different hurdles that immigrants face and to increase the effectiveness of the campaign.

\subsection{Data} \label{sec:data}

We draw our data from administrative sources of the Canton of Zurich. The data includes records of whether and when eligible immigrants submit an application for Swiss citizenship to the City of Zurich during the study period, which allows us to define the outcome variable of our analysis. The data also includes additional covariates which we use to identify and leverage treatment effect heterogeneity. These covariates are age, gender, nationality, years of residency in Switzerland, and years of residency in Zurich. The data also includes an address identifier which allows us to assign the treatment on a building level to minimize contamination by spill-over effects. 

The study sample includes all immigrants in the City of Zurich who satisfy the following criteria:
\begin{enumerate}[noitemsep]
\item They were born on or before June 30, 2003 (i.e., they must have been at least 18 years of age at the start of the study), 
\item they arrived in Switzerland on or before June 30, 2011,
\item they arrived in Zurich City on or before June 30, 2019, 
\item they must have possessed a permanent residence permit (C permit) at the time of randomization (August 2021), and
\item they must not have received any information or encouragement letter in the past.
\end{enumerate}
The first criterion ensures that only adults are in the study. Criteria 2-4 ensure that the entire sample meets the current residency and permit requirements for citizenship. The sample includes 5,145 individuals.

\subsection{Treatment letters}\label{sec:treatments}

Combining insights from the existing literature and our own surveys, we identify three key barriers to naturalization: (i) perceived complexity of the naturalization process, (ii) perceived difficulty of and uncertainty about naturalization requirements and (iii) perception that naturalization is not welcome. 
In collaboration with the City of Zurich, we developed three treatment letters where each letter puts emphasis on one of the hurdles. Each treatment involves the receipt of a letter sent by representatives of the City of Zurich. The treatments differ in the sender, content, wording and design of the letters.The per-unit costs of the three treatments range between 1.20 and 1.50 CHF, and are thus negligible compared to the fiscal benefits of naturalization.\footnote{\citet{hainmueller2019effect} quantify the long-term effect of naturalization on immigrants' earnings at CHF 4,500 per year, which implies an increase in tax revenues for Swiss municipalities of at least CHF 450 per year.} We chose to develop distinct letters to keep the letters brief and understandable, thus avoiding the risk of an informational overload \citep{haaland2023}. The letters, including enclosed flyers, were written in German. Appendix \ref{appendix:letters} contains copies of the original letters in German as well as an English translation. 

The \emph{Complexity letter} consists of a short informational cover letter written by the City Clerk of the City of Zurich (see Appendix \ref{information-letter-appendix}) and a flyer. The half-page cover letter informs recipients that they meet the basic requirements for Swiss citizenship and directs them to sources of further information about the citizenship application process. The flyer included in the \emph{Complexity letter} (shown in Figure~\ref{complexity-flyer-appendix}) attempts to tackle the perceived complexity of the naturalization process. The left-hand side of the flyer shows a video screenshot and a QR code that directs readers to the video, explaining the naturalization process in a simplified way. The right-hand side encourages readers to scan another QR code redirecting to the contact and advice webpage\footnote{The first QR code redirects to \url{https://www.stadt-zuerich.ch/portal/de/index/politik_u_recht/einbuergerungen.html} (last accessed on December 7, 2022). The second QR code redirects to \url{https://www.stadt-zuerich.ch/portal/de/index/politik_u_recht/einbuergerungen/kontakt-und-beratung.html} (last accessed on December 7, 2022).} of the City of Zurich's citizenship office.

The \emph{Requirements letter} includes the same short informational cover letter as the \emph{Complexity letter} but uses a different flyer addressing the perceived difficulty of the naturalization process (see Appendix \ref{requirements-flyer-appendix}). This flyer is also divided into two sections, each containing a descriptive text and a QR code. The QR code on the left-hand side redirects to the targeted, free-of-charge mobile application, which allows immigrants to study for the civics exam and test their knowledge with practice questions.\footnote{The mobile application is developed by the City of Zurich and named \textit{Einbügerungstest Code Schweiz},  which translates to Naturalization Test Code Switzerland.}  The section on the right lists the German language requirements for citizenship and the QR code redirects to a webpage containing more detailed information on the language requirements, exam costs, as well as a link to a practice language exam.\footnote{The website, which the QR code redirected to, moved to \url{https://www.stadt-zuerich.ch/portal/de/index/politik_u_recht/einbuergerungen/kenntnisse/sprachlicheanforderungen.html} on October 21, 2022, due to a mistake by the website maintainers. As a consequence, the QR code broke more than five months after the letter was dispatched to wave~2 participants. We show in Table~\ref{tab:main_results_robustness}, where we only consider the naturalization applications recorded up to five months after letter dispatch, that our main results in Table~\ref{tab:main_results} are not affected by this issue. We thus use, in line with the pre-analysis plan, application outcomes recorded seven months after letter dispatch in the remainder of the study.\label{footnote:error}}

The \emph{Welcome letter} is an information and encouragement letter signed by the Mayor of the City of Zurich. The \emph{Welcome letter} attempts to tackle the hurdle stemming from the perception that naturalization is not welcome \citep{hainmueller_who_2013}. The letter includes only a cover letter (shown in Appendix \ref{welcome-letter-appendix}) that is a little less than one page long and contains three sections. The first section informs recipients that they meet the basic eligibility requirements for Swiss citizenship. The second section encourages them to play an active part in Zurich's political life by becoming a citizen. The last section briefly directs to sources for further information about the citizenship application process and states that the City hopes to see them at the next ceremony for new citizens. Hence, compared to the other two treatment letters, this letter puts more emphasis on the emotional and psychological aspects associated with naturalization and only provides minimal information.

\subsection{Experimental design and estimation methodology}\label{sec:study_design}

This section summarizes the pre-registered experimental design, estimation methodology and evaluation strategy.\footnote{The study was pre-registered at \url{https://osf.io/9wf4t}.} In the exploration phase of the project, we randomly divide the sample of $5{,}145$ eligible immigrants into two groups: Group~A (60\% of the sample) receives one of three treatment letters at random from the City of Zurich in October 2021, while Group~B (40\%) received no letter. The randomization design allocates one of the three treatment letters to individuals in Group A by building address and applied block randomization by nationality groups. The randomization by building address reduces the risk of spill-over effects among eligible immigrants living in the same or neighboring households. The block randomization by nationality group ensures that we have a roughly equal share of nationalities in Group A (including each subgroup receiving different letters) and Group B. We block on nationality groups given the importance of this effect moderator in earlier studies \parencite{ward2019large}. The letters for this first wave were delivered on October 8, 2021.

The first-wage application outcomes enable us to estimate the average treatment effect of treatment letter $d$, i.e., $E[Y_i(d)-Y_i(0)]$, and the conditional average treatment effect $E[Y_i(d)-Y_i(0)\vert X_i]$ where we use $Y_i$ to denote the application outcome recorded at the end of March 2022 and $Y_i(d)$ its potential outcome under treatment $d$. The covariates $X_i$ are country group of nationality, age, gender, years lived in Zurich and years lived in Switzerland, which are constant over the sample period. We employ causal forests due to \citet{wager2018,athey2019c}, a non-parametric method for the estimation of heterogeneous treatment effects relying on random forests.

The main objective, however, is to leverage the first-wave application outcomes $Y_i$ and individual characteristics $X_i$ to fit a multi-action policy tree based on the estimation methodology of \citet{zhou2022} outlined in Section~\ref{sec:policy_learning}. To select the tree depth, we consider a validation exercise: In each iteration, we randomly split the wave-1-data (including untreated) into training and test data with a 60/40 split, and sample from each partition separately with replacement to construct bootstrapped training and validation data sets of sizes $n_1=4{,}871$ and $n_2=1{,}857$. We then fit a policy tree on the bootstrapped training data and estimate the difference in reward between alternative policy rules on the bootstrapped validation data.

In the exploitation phase, we field the fitted policy tree on not-yet-treated individuals in Group B. Specifically, in order to evaluate the performance of the policy rule, we randomly subdivide Group B into two sub-groups, referred to as Group B.1 and Group B.2, and send treatment letters to Group B.1 based on the estimated policy rule while Group B.2 receive a random treatment letter (with one-third probability for each letter). We randomize by building address for the random division into Groups B.1 and B.2, as well as for the randomization of treatments within Group B.2. The City of Zurich delivered the letters for the exploitation phase on May 6, 2022.\footnote{Note that for practical reasons, there was a two-month time gap between measuring the application outcomes in March 2022 and sending out the letter in May.}

The evaluation compares the policy tree against no treatment, random treatment allocation, and conventional one-size-fits-all policy rules that always assign the same treatment to everyone, ignoring treatment effect heterogeneity. To this end, we estimate models of the form:
\begin{equation}
    Y_{it} = W_{it}^\prime\beta + f(X_i,\delta_t) + \varepsilon_{it} \label{eq:main_model}
\end{equation}
where $Y_{it}$ is the application outcome of eligible immigrant $i$ at the end of wave $t\in\{1,2\}$. We add the wave subscript $t$ to accommodate the two-wave structure of the data. The outcomes for the evaluation analysis were recorded approximately 7 months after the date of letter dispatch $t$.\footnote{The application outcomes for the evaluation analysis were recorded in May 9 and December 9, 2022, respectively. In Table~\ref{tab:main_results_robustness}, we provide alternative results where we consider all application outcomes until March 21 and October 21, 2022, respectively (see fn.~\ref{footnote:error}).} The time-invariant covariates $X_i$ are defined above. $\delta_t$ is a dummy for wave $t\in\{1,2\}$, and accounts for seasonal effects and other external shocks that may affect application rates. The vector $W_{it}$ assigns individuals to treatment groups, and is defined as $W_{it}=(\textit{Letter}^1_{it},\textit{Letter}^2_{it},\textit{Letter}_{it}^3,\textit{Nothing}_{it},\linebreak[1] \textit{PolicyTree}_{it})$ or $W_{it}=\left(\textit{Random}_{it},\textit{Nothing}_{it},\textit{PolicyTree}_{it}\right)$, respectively, 
where $\textit{Letter}^j_{it}$ is set to 1 if the individual $i$ was randomly assigned to treatment letter $j\in\{1,2,3\}$ for wave $t$, 0 otherwise. $\textit{Nothing}_{it}$ is set to 1 if the individual $i$ has received no treatment in wave $t$, and $\textit{PolicyTree}_{it}$ equals 1 if individual $i$ has received the treatment letter assigned to them by the policy tree. Finally, $\textit{Random}_{it}$ is set to 1 if individual $i$ was randomly assigned to one of the three letters, 0 otherwise.

We estimate \eqref{eq:main_model} by linear regression using only the elementary controls, but also consider more flexible methods. Namely, we use Post-Double Selection Lasso \citep[PDS-Lasso;][]{belloni2014inference} and Double-Debiased Machine Learning \citep[DDML;][]{chernozhukov2018b} where we extend the set of controls by interaction terms and second-order polynomials.\footnote{For the Post-Double Selection Lasso, we use cluster-robust penalty loadings of \citet{belloni2016a}. With regard to DDML, we use 10 cross-fitting folds, 5 cross-fitting repetitions and use stacking with a set of candidate learners including linear regression, lasso, ridge, random forests and gradient boosting \citep{Ahrens2023_stacking}}. We cluster standard errors by building addresses, i.e., the level at which the treatment was applied.\footnote{We note that the clustered standard errors do not account for sampling variability arising from the estimation of the policy rules. The issue is akin to the well-known generated regressor problem which occurs when a regressor is unobserved and replaced by a first-step estimate. The generated regressor problem is usually addressed using standard-error adjustments \citet{pagan1984,murphy1985} or, most commonly, using bootstrapping; see e.g.\ review in \citet{chen2023} or \citet{wooldridge2010econometric}. Neither of these approaches is feasible in our setting. Analytical standard errors are, to our knowledge, not available for this specific problem. Bootstrapping or other resampling techniques would require us to repeatedly field policy rules fitted on bootstrapped samples of the data in order to capture the variability in estimated policy rules, which is practically infeasible. We thus interpret the standard errors with caution.}

\subsection{Results from the exploration phase: Learning the policy rule}\label{sec:wave_1_results}

We begin by analyzing the results from the exploration phase of the experiment using naturalization applications received by the end of March 2022 (i.e., wave 1). Descriptive statistics of the wave-1-data are provided in Table~\ref{tab:wave1}. We proceed in three steps: estimation of (conditional) averages of treatment effects, tuning policy trees using a validation exercise and fitting the policy tree on the full wave-1-data.

\begin{table}[ht]
    \centering\footnotesize
    \begin{tabularx}{.75\linewidth}{lXrrrrr}
      \toprule\midrule
     &  & \it Avg. & \it St.dev. & \it Min & \it Max & \it Obs. \\ 
      \midrule
      \multicolumn{2}{l}{\it Dependent variable:} \\
        \partialinput{11}{11}{tables/descriptives_wave1.tex}
        \multicolumn{2}{l}{\it Covariates:} \\
        \partialinput{9}{10}{tables/descriptives_wave1.tex}
        \partialinput{12}{13}{tables/descriptives_wave1.tex}
        \multicolumn{2}{l}{\it Regions:} \\
        \partialinput{14}{24}{tables/descriptives_wave1.tex}
       \midrule\bottomrule
       \multicolumn{7}{p{.73\linewidth}}{{\it Notes:} The table shows summary statistics for covariates and dependent variables measured until March 2022.}\\
    \end{tabularx}
    \caption{Descriptive statistics of wave-1 data}\label{tab:wave1}
\end{table}

\begin{figure}[htb]
    \centering\singlespacing
    \includegraphics[width=\linewidth]{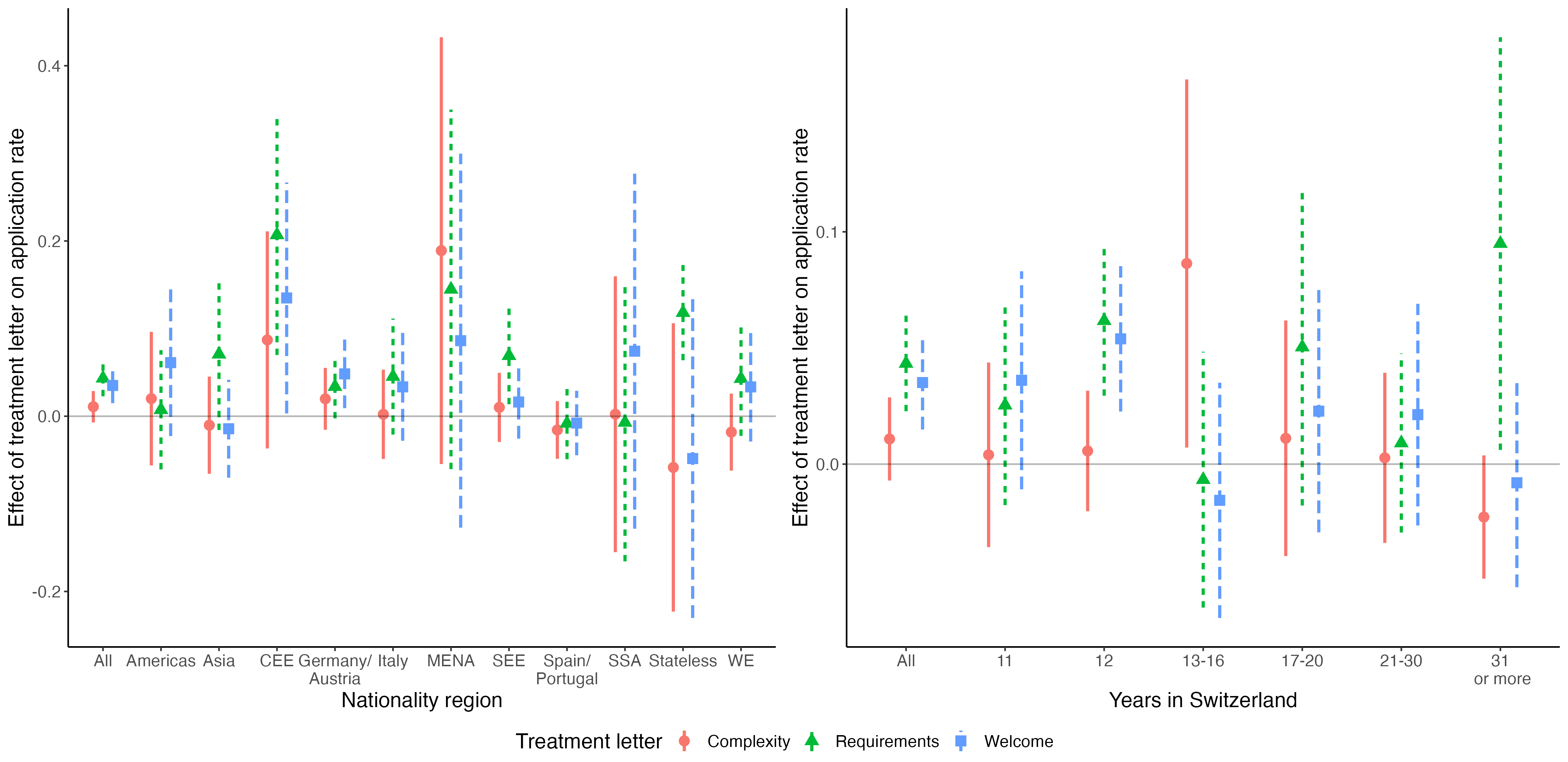}
    \par\medskip
    \parbox{\linewidth}{\footnotesize\emph{Notes:} The figures shows the average and conditional average treatment effects by group where the groups are formed based on nationality and years of residence in Switzerland. The regions are the Americas, Asia, Central and East Europe (CEE), Germany and Austria, Italy, Middle East and Northern Africa (MENA), South-East Europe, Spain and Portugal, Stateless and Sub-Saharian Africa (SSA). The treatment effects are estimated using a multi-arm causal forest and using the R package \texttt{grf} \citep{wager2018,athey2019c,grf}.}\\[.5cm]
    \caption{Average and conditional average treatment effects}
    \label{fig:treatment_effects}
\end{figure}

First, we fit a multi-arm causal forest to estimate average treatment effects, as well as conditional average treatment effects by nationality group and years lived in Switzerland \citep{wager2018,athey2019c}. Results are displayed in Figure~\ref{fig:treatment_effects}.\footnote{We removed 274 individuals who moved between October 2021 and March 2022, resulting in an estimation sample of $4{,}871$ individuals.} The average treatment effects for the first-wave sample imply that the \emph{Complexity letter} increases application rates by $1.08$ p.p.\ ($s.e.{=}0.91$), the \emph{Requirements letter} by $4.33$ p.p.\ ($s.e.{=}1.04$), and the \emph{Welcome letter}  by $3.51$ p.p.\ ($s.e.{=}1.03$), relative to the control condition of no letter.\footnote{See \citet{hangartner2023citizenship} for a discussion the letters' efficacy in overcoming specific hurdles.} 

The left panel of Figure~\ref{fig:treatment_effects} shows only moderate heterogeneity in treatment effects by nationality. The \emph{Welcome letter} appears to have slightly stronger effects for immigrants from Germany and Austria, consistent with the idea that Germans and Austrians do not perceive complexity and difficulty as major hurdles due to their cultural proximity and language. At the same time, the \emph{Welcome letter} is also the most effective letter for immigrants from the Americas, which could indicate that this minority group does not feel very welcome in Switzerland. The relative effect size of the \emph{Requirements letter} is particularly large for immigrants from Central-Eastern and South-Eastern Europe, as well as for stateless immigrants. The right panel of Figure~\ref{fig:treatment_effects} indicates that the \emph{Complexity letter} has the largest effect on application rates among eligible immigrants who have lived between 13 and 16 years in Switzerland. In contrast, eligible immigrants who have lived for more than 30 years in Switzerland are especially receptive to the requirements letter, suggesting that the perceived difficulty of the naturalization process may discourage some eligible immigrants from applying over long periods. This effect may also be partially driven by age since we also find the \emph{Requirements letter} to have the largest effect among immigrants aged 46 and above (see Figure~\ref{fig:fig:treatment_effects_appendix} in the Appendix). Finally, we find that men are slightly more receptive to the letter treatments overall than women, but the ranking of treatment letter efficacy is the same (see Figure~\ref{fig:fig:treatment_effects_appendix}). 

Second, we conduct the validation exercise outlined above to assess the out-of-sample performance of various policy rules and to select the tree depth of the policy tree.  We focus on policy trees with tree depths of 2 and 3, as well as a hybrid policy tree of depth 4. For comparison, we consider (i) one-size-fits-all rules that always assign one of the \emph{Complexity}, \emph{Requirements} or \emph{Welcome} letters, (ii) random allocation of one of the three letters, and (iii) a model-free plug-in rule that assigns the treatment for which the estimated reward is the largest. We repeat the exercise 500 times and report average differences in rewards and bootstrapped standard errors in Table~\ref{tab:simulation_exercise}.\footnote{We opted for this approach rather than $K$-fold cross-validation as it allows us to match the sizes of the training and validation data to the actual sample sizes. However, we obtain similar results when applying $K$-fold cross-validation.} The table reports in each column the gain in reward of a specific policy choice compared to alternative policy rules (shown in rows). For instance, the coefficient of 1.026 ($s.e.=.99$) in the top-left entry corresponds to the gain in reward of a one-size-fits-all policy rule assigning the \emph{Complexity letter} to everyone relative to a policy rule assigning no letter. We find that all three policy trees outperform each individual treatment letter as well as random treatment allocation. Among the three policy trees, the tree of depth 3 performs marginally better than trees of depth 2 and 4. As expected, the plug-in rule shows overall the best performance. However, the plug-in rule provides no insights into the drivers of treatment effects. The results thus highlight the trade-off between interpretability and performance but also show that, in this context, the best-performing policy tree is able to reach more than 85\% of the performance of the plug-in rule.

\begin{table}[htb]
    \scriptsize\singlespacing\centering
    \newcommand{\starsss}{$^{***}$}
    \newcommand{\starss}{$^{**}$}
    \newcommand{\stars}{$^{*}$}
    \caption{The effect of the policy rule compared randomization, always the same treatment and no treatment}\label{tab:simulation_exercise}
    \medskip
        \def\sym#1{\ifmmode^{#1}\else\(^{#1}\)\fi}
        \begin{tabular}{l*{10}{c}} \hline\hline
         &  \multicolumn{3}{c}{\it One-size-fits-all}  & {\it Random}  & \multicolumn{3}{c}{\it Policy tree} & {\it Plug-in} \\
             &  {\it Complexity} &   {\it Requirem.} &  {\it Welcome} & {\it treatment} &   $d=2$ &  $d=3$ &   $d=4$ & {\it rule}  \\ 
           \hline
        \partialinput{5}{20}{tables/comparison_table.tex}
        \hline\hline
       \end{tabular}
      \par\medskip
      \parbox{\linewidth}{
      \textit{Notes:} The table reports the difference in estimated rewards between policy rules based on wave-1 data (including untreated immigrants of Group B). Specifically, each cell corresponds the gain in reward of a specific policy rule (shown in columns) relative to alternative policy rules (listed in rows). The results are based on a resampling exercise where we randomly split the wave-1 data into training and test data using a 60/40 split, and separately draw $n_1=4871$ and $n_2=1857$ observations with replacement from the training and test data. We use 500 repetitions and report the average difference in rewards and associated bootstrapped standard errors. Significance levels: ${}^{\star\star\star}\, 0.01, {}^{\star\star}\,  0.05, {}^{\star}\, 0.1$.}
\end{table}

\begin{figure}[htb]
    \centering\singlespacing
    \includegraphics[width=.9\linewidth]{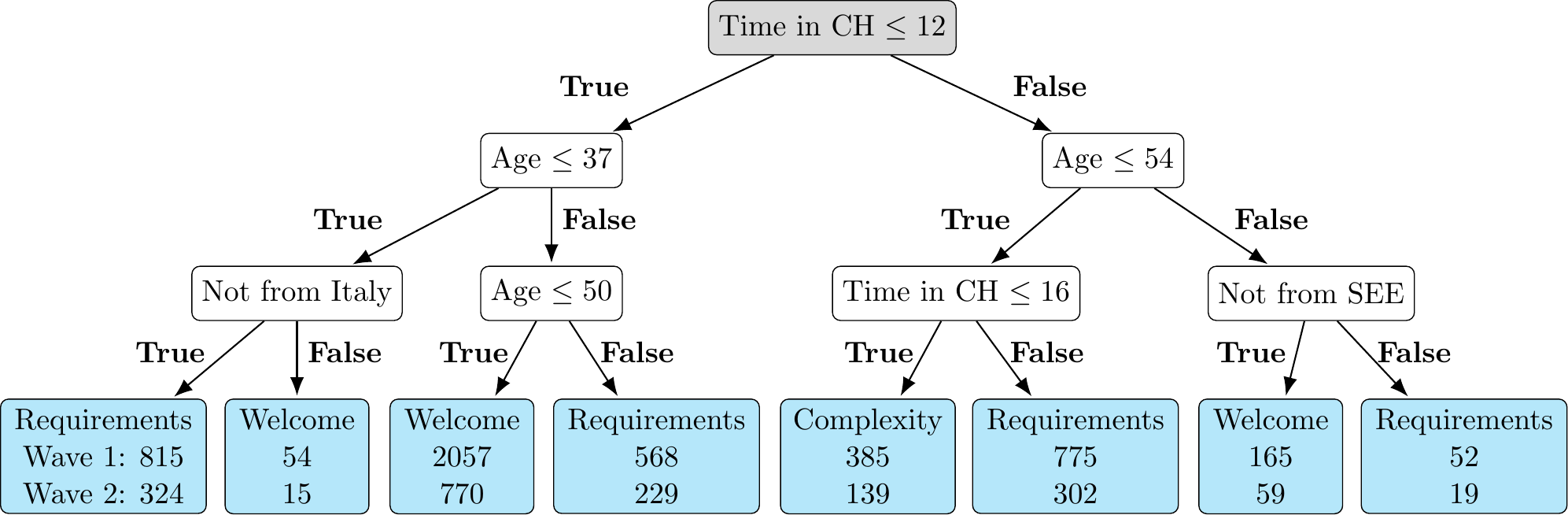}\\[.4cm]
    \parbox{\linewidth}{\footnotesize\emph{Notes:} The figure shows the policy tree fitted to data from wave 1. The size of the training sample is $4{,}871$. The numbers at the bottom indicate the number of individuals assigned to each terminal node in the training sample and in Group B.}\\[.5cm]
    \caption{Fitted policy tree}
    \label{fig:policy_tree}
\end{figure}

Third, in light of the advantages and limited costs of policy trees in this setting, we opted for implementing the policy tree of depth 3. Following the approach of \citet{zhou2022} as outlined in Section~\ref{sec:policy_learning}, we trained the policy tree on wave 1 data, including Group A (who received a letter in the first wave) and Group B (who did not receive a letter in the first wave). Since we randomized treatment assignment in the first wave, we did not need to estimate the propensity scores but plugged the known treatment shares into \eqref{eq:caipwl}.\footnote{We note that in a setting where $e_a(X_i)$ is known, the IPW estimator of the reward in \eqref{eq:qipw} is also applicable. We find in simulations that the AIPW estimator using the known propensity scores outperforms the IPW estimator.} We used multi-arm causal forests to estimate the double robust scores, although other estimators are possible. The fitted policy tree $\hat\pi$ of depth three is displayed in Figure~\ref{fig:policy_tree}. The boxes at the bottom of the tree show the assigned treatment for the wave-1 sample and the wave-2 sample (i.e., Group B) per terminal node. For instance, the very-left branch assigns individuals who have spent no more than 12 years in Switzerland, are aged 37 years or younger, and who are not from Italy to the requirements treatment. 815 individuals in total and 324 individuals from Group B fall into that category. In total, 139 individuals of Group B are assigned to the \emph{Complexity letter}, 874 individuals to the \emph{Requirements letter} and 844 to the \emph{Welcome letter}.\footnote{We assigned policies for Groups B.1 and B.2 after removing individuals who either applied without being treated (99 individuals) or moved out of the municipality of Zurich (101 individuals).} The splits in the tree are based on years in Switzerland, age, and only two nationality indicators, but no split is based on gender confirming that the relative performance of each letter is the same for women and men. It is also noteworthy that no individuals were assigned to receive no letter, which suggests that at least one of the three letters has a positive effect for every individual. 

\subsection{Results from the exploitation phase: Evaluating the policy rule}\label{sec:wave_2_results}

Table~\ref{tab:main_results} shows the results of the evaluation based on estimating versions of \eqref{eq:main_model} using OLS (columns 1-3), PDS lasso (col.\ 4-5), and DDML (col.\ 6-7). The sample includes only wave 2 in column 1, and both waves in the remaining columns. The reference group in column 1 is random treatment allocation, while the base group in columns 2-7 is no treatment. Panel A reports the coefficient estimates and Panel B compares the policy rule using policy trees against each individual treatment letter and random treatment allocation.

\begin{table}[htb]
    \scriptsize\singlespacing\centering
    \caption{The effect of the policy rule compared randomization, always the same treatment and no treatment}\label{tab:main_results}
    \medskip
        \def\sym#1{\ifmmode^{#1}\else\(^{#1}\)\fi}
        \begin{tabular}{l*{8}{c}}
        \hline\hline
            &\multicolumn{1}{c}{(1)}&\multicolumn{1}{c}{(2)}&\multicolumn{1}{c}{(3)}&\multicolumn{1}{c}{(4)}&\multicolumn{1}{c}{(5)}&\multicolumn{1}{c}{(6)} &\multicolumn{1}{c}{(7)} \\
            &\multicolumn{7}{c}{\it Dependent variable: Naturalization application}\\
        \hline
        \multicolumn{7}{l}{\it Panel A. Coefficient estimates} \\
        \partialinput{7}{20}{tables/main_output.tex} \\
        \hline 

        \hline
        \multicolumn{7}{l}{\it Panel B. Comparison of Policy tree with:}\\
        \partialinput{22}{29}{tables/main_output.tex}
        \hline
        Sample & Wave 2 & Wave 1-2 & Wave 1-2 & Wave 1-2 & Wave 1-2 & Wave 1-2 & Wave 1-2 \\
        \partialinput{30}{32}{tables/main_output.tex}
        \hline\hline
       \end{tabular}
      \par\medskip
      \parbox{\linewidth}{
      \textit{Notes:} The table reports results from estimating versions of \eqref{eq:main_model} using OLS (columns 1-3), PDS-Lasso (columns 4-5) and DDML (columns 6-7). Column 1 only uses data from wave 2; the remaining columns use the full data set. The reference group in column 1 is random treatment allocation; no treatment in columns 2-7. Panel A reports the coefficient estimates. Panel B compares the policy rule using policy trees against always assigning the same treatment to everyone and random treatment allocation. Covariates include the region of nationality, age, gender, years lived in Zurich and years lived in Switzerland. \\ 
      Standard errors are clustered at building address level. 
      \sym{*} \(p<0.05\), \sym{**} \(p<0.01\), \sym{***} \(p<0.001\)
      }
\end{table}

According to the OLS results in columns 1-3, the treatment assignment by policy tree increased the application rate by $1.79$ ($s.e.{=}1.36$) to $1.90$ p.p.\ ($1.36$) relative to random treatment, and by around $5.13$ p.p.\ ($1.61$) compared to no treatment. Random allocation is associated with an application rate increase of approximately $3.23$ p.p.\ ($0.82$).  Turning to the individual treatments, we find that the \emph{Welcome letter} yields overall the largest increase in application take-up with an effect size around $3.79$ p.p.\ ($1.07$), closely followed by the \emph{Requirements letter} with an effect size around $3.65$ p.p.\ ($1.10$). The \emph{Complexity letter} performs substantially worse in comparison, with an effect size of   $2.23$ ($s.e.{=}1.04$). Panel B shows that the policy tree performs better than random treatment or each individual treatment option. The take-up increase compared to the best-performing individual treatment (the \emph{Welcome letter}) is 1.03 p.p.\ but statistically insignificant. The PDS lasso estimates are almost identical and the DDML estimator yields effect sizes only marginally smaller.\footnote{\revised{Appendix Table~\ref{tab:logit_robustness} also shows alternative results using logistic regression. The average marginal effects from logistic regression are almost identical to those from OLS.}}

\section{Conclusion}\label{sec:conclusion}

This paper employs policy trees for assigning eligible immigrants to the information and encouragement treatment that is most likely to address hurdles on their path to citizenship and boost their propensity to naturalize. We evaluate the benefits of this policy rule using a tailored two-phase field experiment. During the exploration phase, we randomly assign eligible immigrants to one of three treatment arms or the control group, based on which we estimate average treatment effects and train the policy tree. We find that despite its simplicity, the optimal policy tree of depth 3 captures more than 85\% of the treatment effect heterogeneity (relative to a model-free plug-in rule). Next, we move on to the exploitation phase, in which we assign the subjects that belonged to the control group in the previous phase to either the policy tree or randomly to one of the three treatments. We find that the policy tree outperforms the best-performing individual treatment slightly. While these differences are not statistically significant, it is worth noting that these benefits persist in a context with at most moderate levels of treatment effect heterogeneity and come at little additional costs. 

Policy trees possess several advantages that make them particularly suited for policymakers and researchers interested in tailoring treatment assignment to the specific needs of increasingly diverse populations. Policy trees are transparent in terms of which variables guide treatment assignment, they are simple to visualize, and intuitive to communicate even to users of the research who lack statistical training. While using machine learning to personalize treatment assignments raises a host of important ethical and policy questions, we should keep in mind that a one-size-fits-all approach can often exacerbate existing inequalities. For instance, an earlier information letter sent out by the City of Zurich had by far the strongest effects among newly eligible immigrants, which often score higher on multiple integration dimensions compared to more marginalized immigrants who have been residing in the host country for decades without naturalizing \citep{ward2019large}. For all these reasons, we believe that policy trees are a well-suited approach to leverage the potential of tailored treatment assignment in a world where rich background characteristics are increasingly available.




\newpage
\printbibliography

\clearpage

\setcounter{page}{1}


\renewcommand\thetable{\Alph{section}.\arabic{table}} 
\renewcommand\thefigure{\Alph{section}.\arabic{figure}} 
\renewcommand\thesection{\Alph{section}} 
\setcounter{section}{0}

\section{Supplementary materials}
\label{materials-appendix}

\subsection{Additional results on heterogeneous treatment effects}

 \begin{figure}[H]
    \centering\singlespacing
    \begin{subfigure}{.48\linewidth}
    \includegraphics[width=\linewidth]{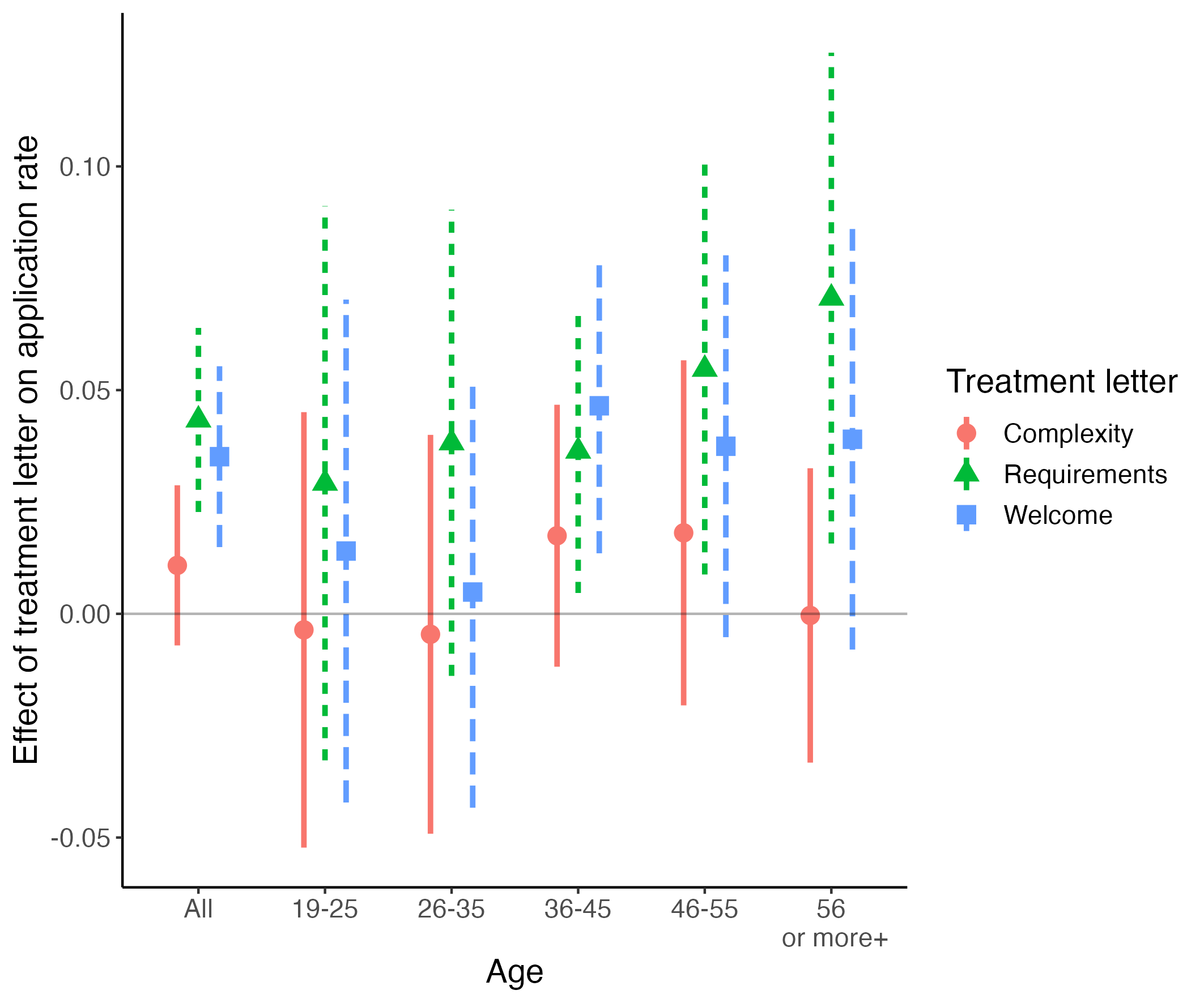}
    \end{subfigure}
    \begin{subfigure}{.48\linewidth}
    \includegraphics[width=\linewidth]{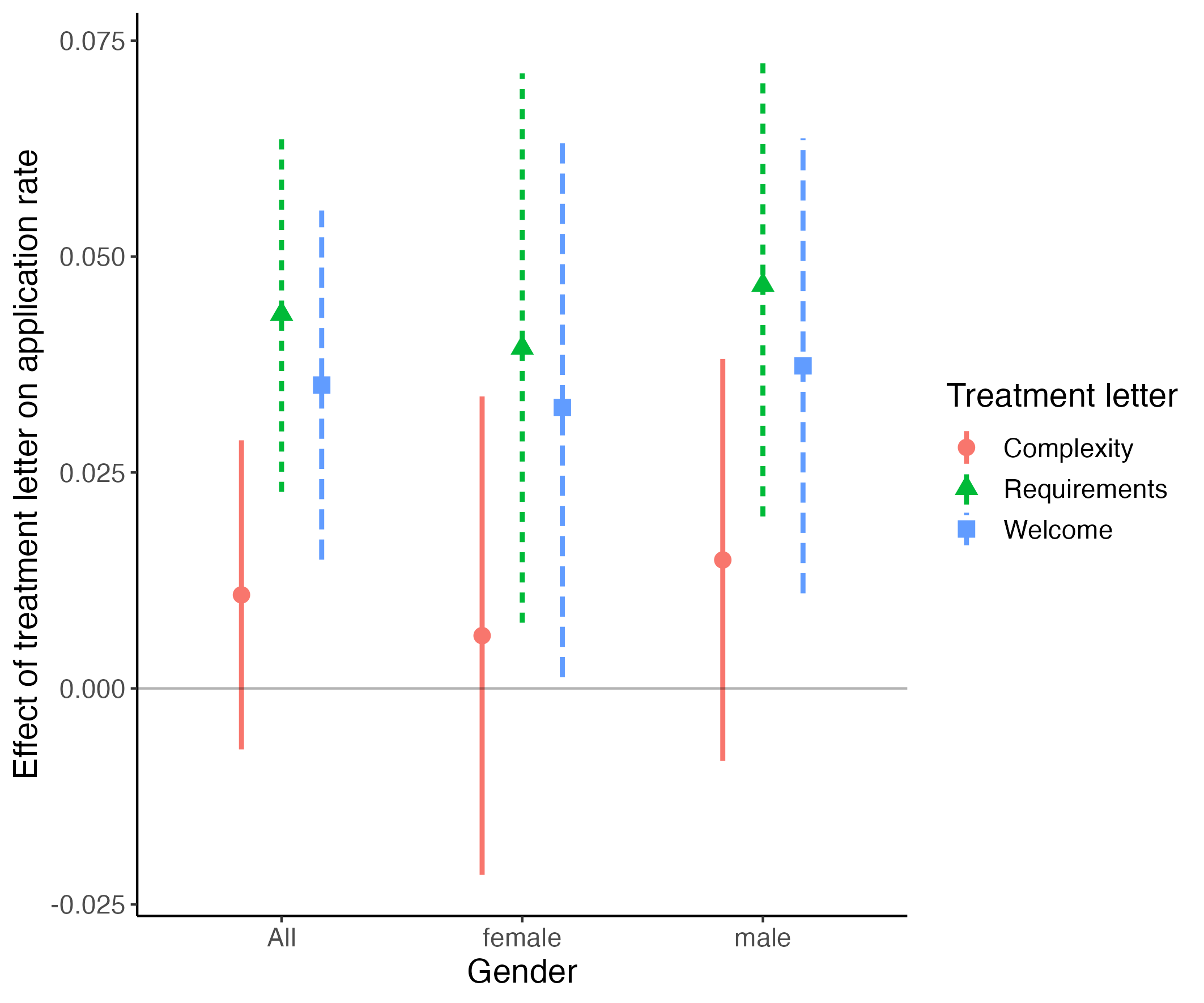}
    \end{subfigure}
    \par\medskip
    \parbox{\linewidth}{\footnotesize\emph{Notes:} The figure shows the policy tree fitted to data from wave 1. The sample size is $4,871$. The numbers at the bottom indicate the number of individuals assigned to each terminal node in wave 1 (in-sample) and wave 2 (out-of-sample). In total, 139 individuals are assigned to the complexity letter, 874 individuals to the \emph{Requirements letter} and 844 to the welcome letter. The estimation was implemented using the R packages \texttt{grf} and \texttt{policytree} \citep{grf,policytree}.}\\[.5cm]
    \caption{Policy tree of depth 3}
    \label{fig:fig:treatment_effects_appendix}
\end{figure}


\newpage\subsection{Letters and flyers sent by the City of Zurich}\label{appendix:letters}

\subsubsection{Informational cover letter}
\label{information-letter-appendix}

\begin{figure}[H]
\centering
\fbox{
\includegraphics[width = .75\textwidth ]{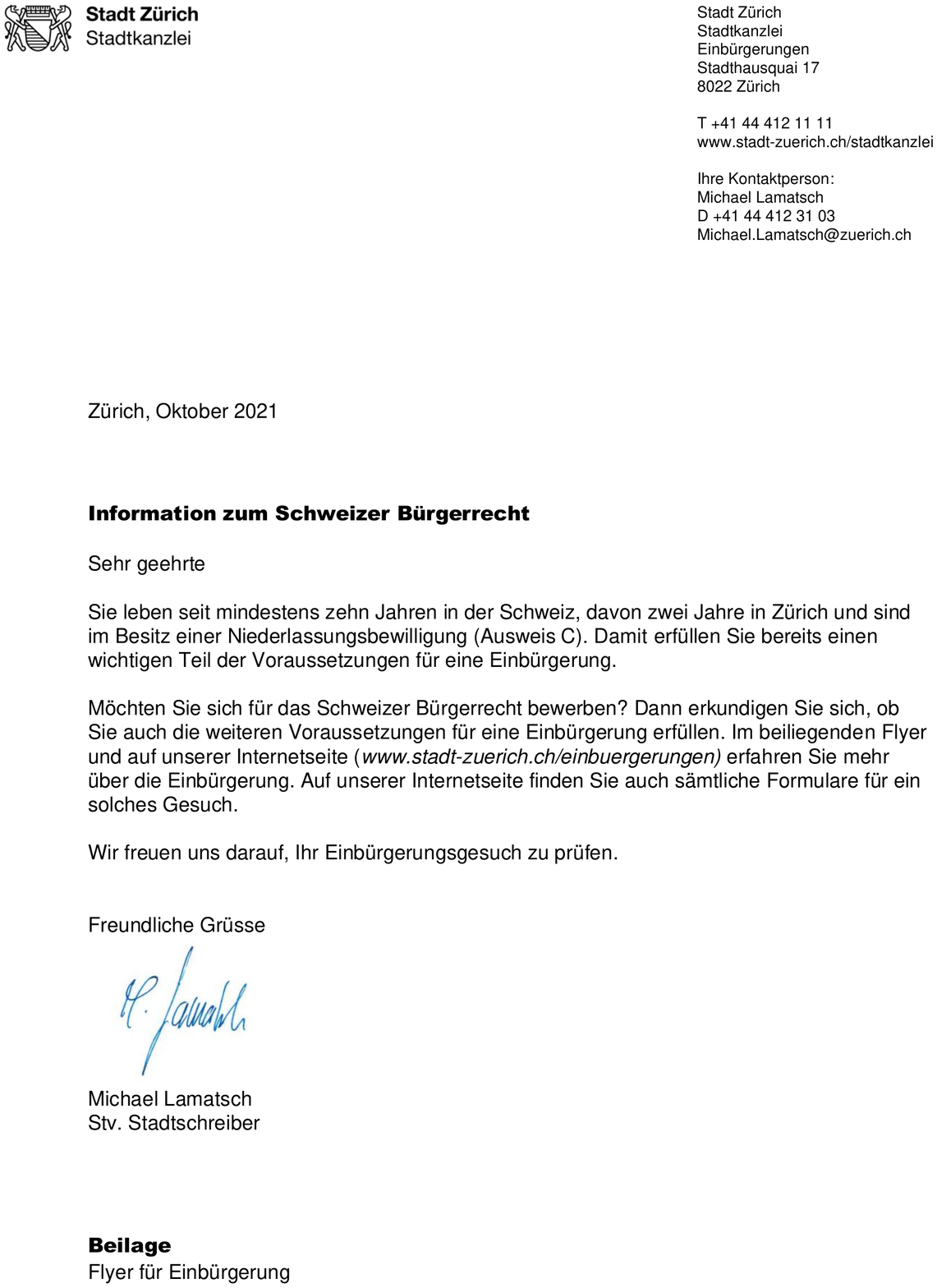}
}\par\vspace{5pt}
\caption{Information Cover Letter. Original in German.}
\end{figure}

\clearpage

\begin{figure}[h!]
\begin{center}\singlespacing 
\begin{tcolorbox}[colback=white,width=.9\linewidth]
\noindent Zurich, October 2021 \\

\noindent \textbf{Important information regarding Swiss citizenship} \\

\noindent Dear \\

\noindent You have been living in Switzerland for at least ten years, two of which in Zurich, and  are in possession of a permanent residence permit (Permit C). You thus already fulfill an important requirement to become a Swiss citizen. \\

\noindent \textbf{Take an active part in democracy} \\
\noindent More than 420,000 people live in Zurich, including 130,000 foreigners from all continents. Foreigners are an important part of our society and contribute and contribute to a good and attractive community in our city. They all make the city of Zurich a successful and livable city. \\

\noindent If you are planning your future in Switzerland, it is important that you can also shape our common future and that you can vote and elect. Because a democracy is only alive and strong if as many people as possible have a say in politics. Thanks to the Swiss citizenship, you have this opportunity to have a voice in politics. That is why the City Council is committed to ensuring that people who meet the requirements for citizenship also naturalize. \\

\noindent \textbf{Inform yourself about naturalization} \\
\noindent Find out whether you meet the other requirements for naturalization. You can find out more about naturalization in the information sheet (enclosure) or on our website (www.stadt-zuerich.ch/einbuergerungen). There you will also find all the forms for a naturalization application. \\

\noindent Every year, the City Council welcomes the new Swiss citizens of the city of Zurich at a ceremony. We would be pleased to also welcome you as a Swiss citizen soon. \\

\noindent Kind regards \\

\noindent Corine Mauch \\
\noindent Mayor \\

\noindent Dr. Claudia Cuche-Curti \\
\noindent City Clerk \\

\end{tcolorbox}
\end{center}
\caption{Informational cover letter. English translation.}
\end{figure}

\clearpage

\subsubsection{Flyer enclosed in complexity letter}
\label{complexity-flyer-appendix}

\begin{figure}[H]
\centering
\begin{subfigure}{\linewidth}\centering
   \includegraphics[width = .8\textwidth, height = !]{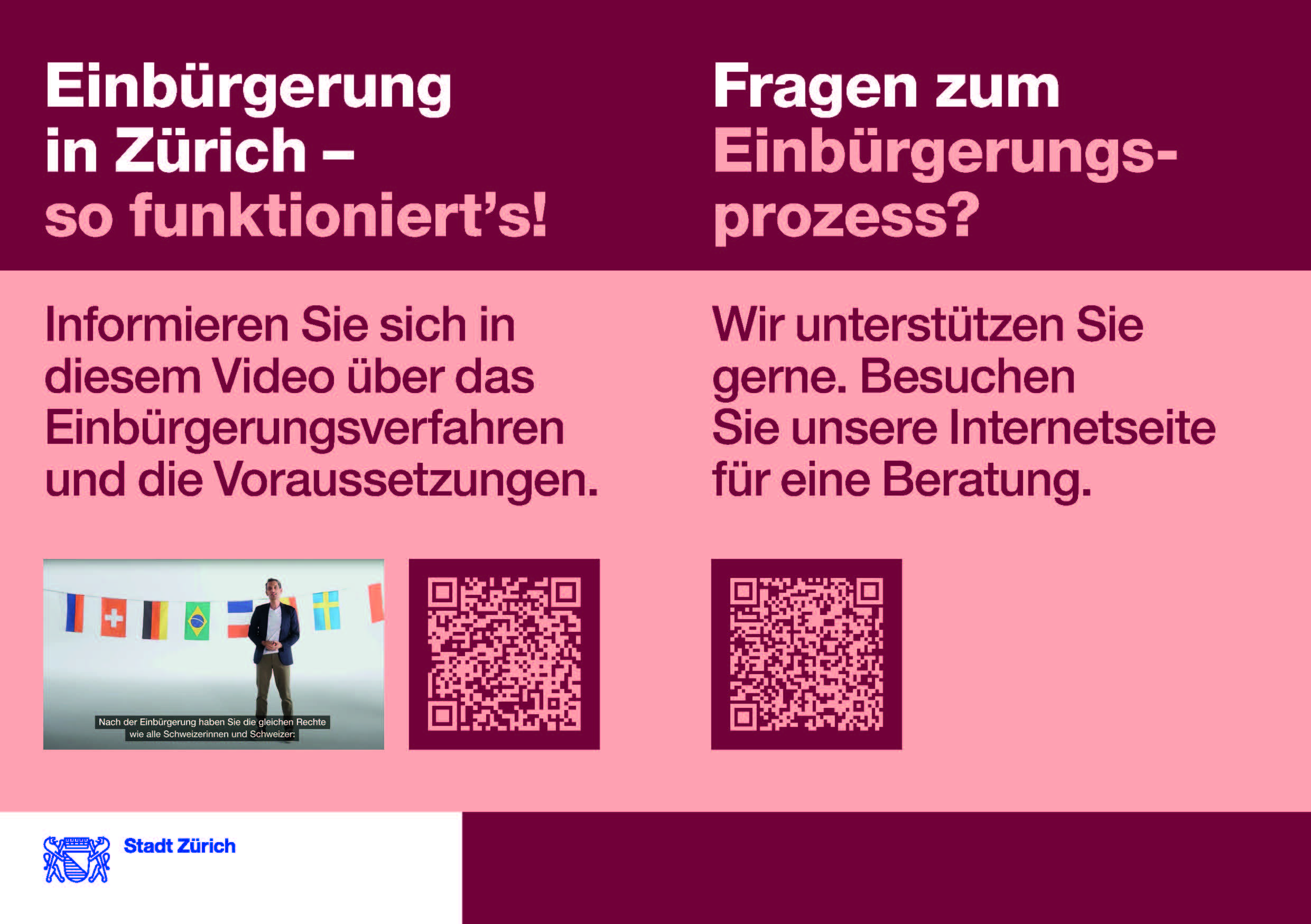} 
   \caption{Original German language version}
\end{subfigure}\\\medskip
\begin{subfigure}{\linewidth}\centering
\includegraphics[width = .8\textwidth, height = !]{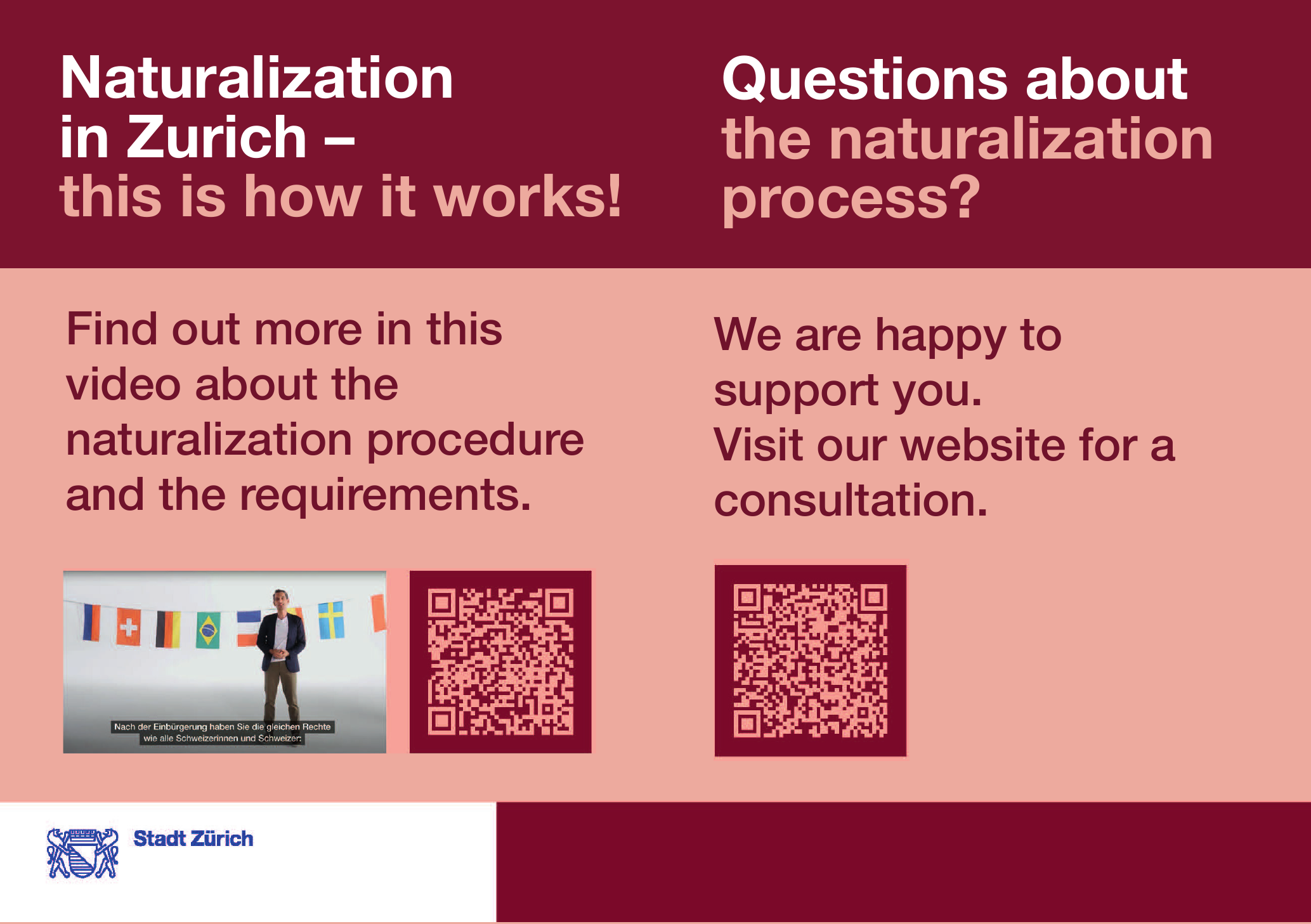}
    \caption{English translation}
\end{subfigure}
\smallskip
\caption{Flyer enclosed in complexity letter}
\end{figure}

\clearpage

\subsubsection{Flyer enclosed in requirements letter}
\label{requirements-flyer-appendix}

\begin{figure}[H]
\centering
\begin{subfigure}{\linewidth}\centering
   \includegraphics[width = .8\textwidth, height = !]{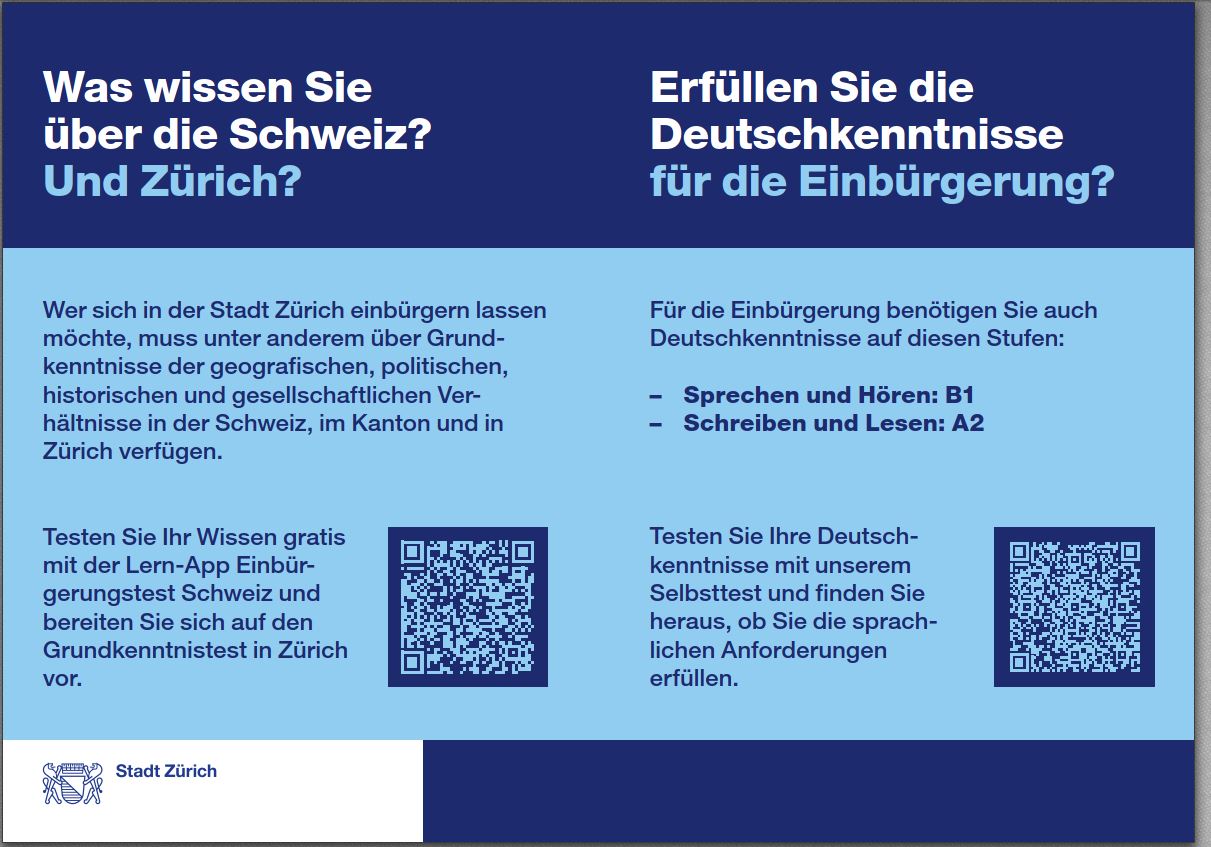} 
   \caption{Original German language version}
\end{subfigure}\\\medskip
\begin{subfigure}{\linewidth}\centering
\includegraphics[width = .8\textwidth, height = !]{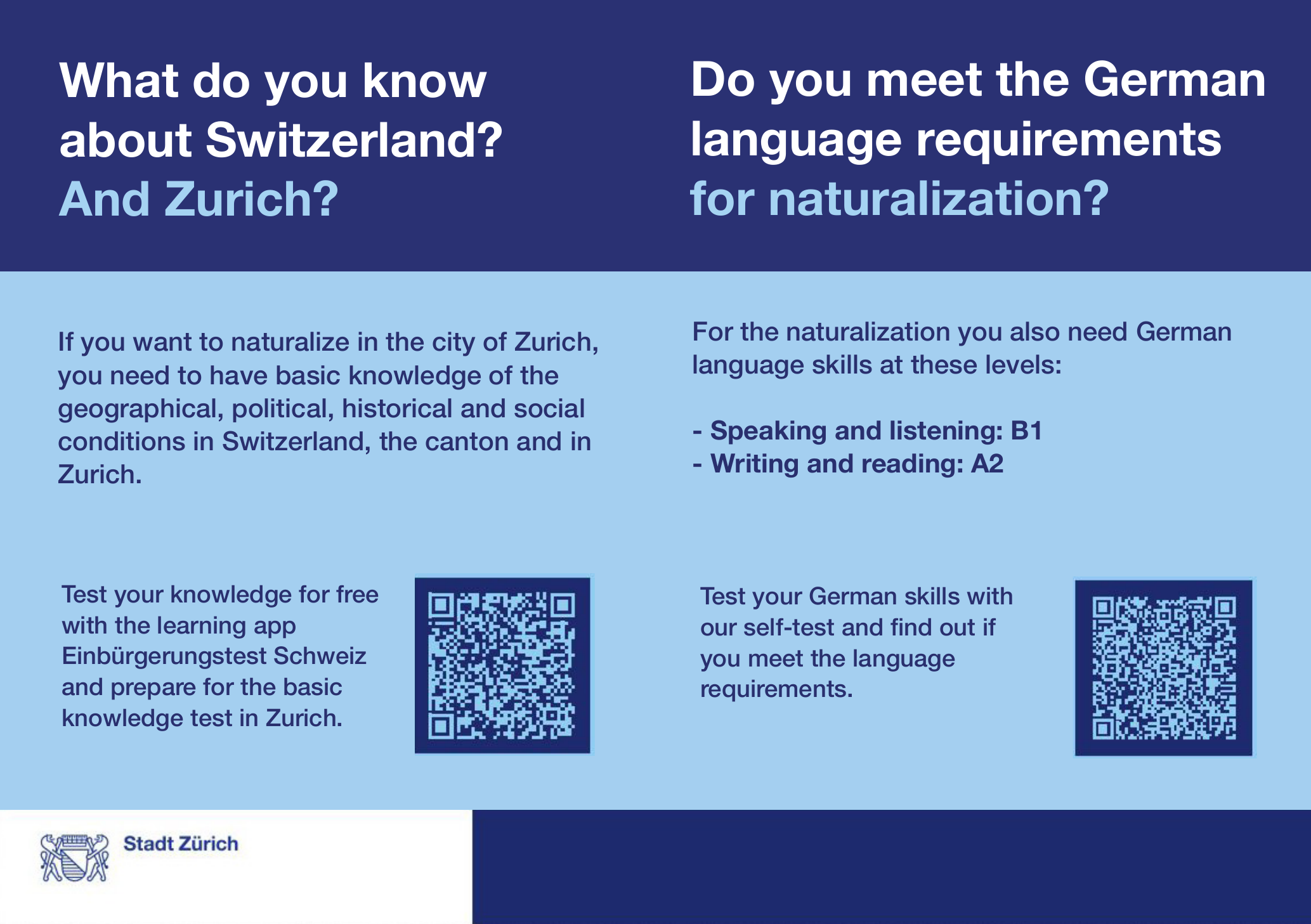}
    \caption{English translation}
\end{subfigure}
\vspace{5pt}
\caption{Flyer enclosed in requirements letter}
\end{figure}

\clearpage
\subsubsection{Welcome cover letter}
\label{welcome-letter-appendix}

\begin{figure}[H]
\centering
\fbox{
\includegraphics[width = .8\textwidth]{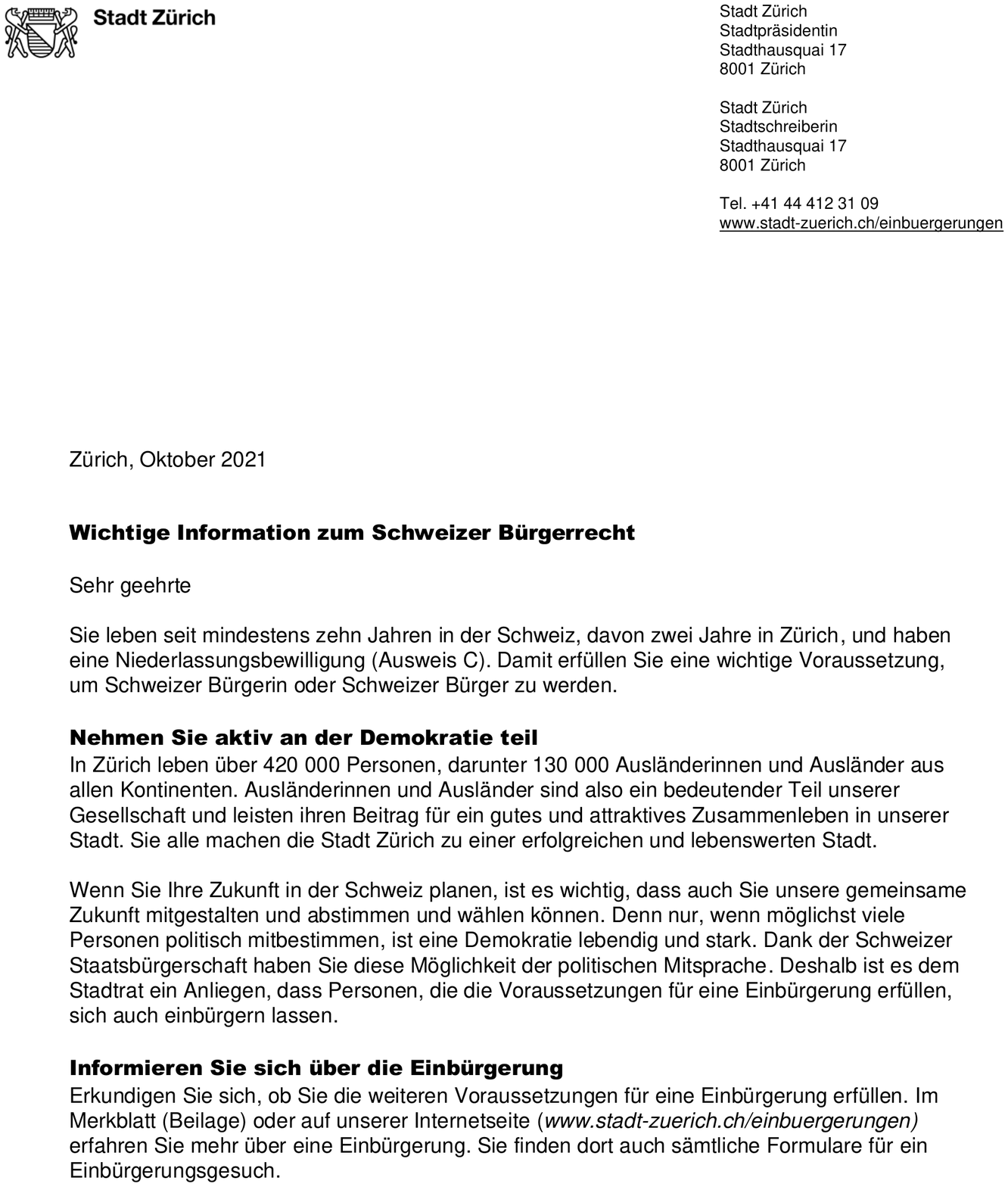}
}\par\vspace{5pt}
\caption{Welcome cover letter. Original in German}
\end{figure}

\begin{figure}[H]
\begin{center}\singlespacing 
\begin{tcolorbox}[colback=white,width=.9\linewidth]
\noindent Zurich, October 2021 \\ 

\noindent \textbf{Information on Swiss citizenship } \\

\noindent Dear \\

\noindent You have been living in Switzerland for at least ten years, two of which in Zurich, and are in possession of a permanent residence permit (Permit C). You thus already fulfill an important requirement to become a Swiss citizen. \\

\noindent Would you like to apply for Swiss citizenship? Then find out whether you also meet the other requirements for naturalization. You can find out more about naturalization in the enclosed flyer and on our website (www.stadt-zuerich.ch/einbuergerungen). On our website, you will also find all the forms for such an application. \\

\noindent We look forward to reviewing your naturalization application. \\ 

\noindent Kind regards, \\
 
\noindent Michael Lamatsch \\
\noindent Deputy City Clerk \\

\end{tcolorbox}
\end{center}
\caption{Welcome cover letter. English translation}
\end{figure}

\newpage\subsection{Alternative results}

\begin{table}[H]
    \scriptsize\singlespacing\centering
    \caption{The effect of the policy rule compared randomization, always the same treatment and no treatment (alternative results)}\label{tab:main_results_robustness}
    \medskip
        \def\sym#1{\ifmmode^{#1}\else\(^{#1}\)\fi}
        \begin{tabular}{l*{8}{c}}
        \hline\hline
            &\multicolumn{1}{c}{(1)}&\multicolumn{1}{c}{(2)}&\multicolumn{1}{c}{(3)}&\multicolumn{1}{c}{(4)}&\multicolumn{1}{c}{(5)}&\multicolumn{1}{c}{(6)} &\multicolumn{1}{c}{(7)} \\
            &\multicolumn{7}{c}{\it Dependent variable: Naturalization application}\\
        \hline
        \multicolumn{7}{l}{\it Panel A. Coefficient estimates} \\
        \partialinput{7}{20}{tables/main_output_robustness.tex} \\
        \hline 

        \hline
        \multicolumn{7}{l}{\it Panel B. Comparison of Policy tree with:}\\
        \partialinput{22}{29}{tables/main_output_robustness.tex}
        \hline
        Sample & Wave 2 & Wave 1-2 & Wave 1-2 & Wave 1-2 & Wave 1-2 & Wave 1-2 & Wave 1-2 \\
        \partialinput{30}{32}{tables/main_output_robustness.tex}
        \hline\hline
       \end{tabular}
      \par\medskip
      \parbox{\linewidth}{
      \textit{Notes:} In this table, we provide alternative results where we consider only application outcomes recorded until approximately five months after letter dispatch, i.e., until March 21 (for wave 1) and October 21, 2022 (for wave 2), respectively. We provide these alternative results to verify the robustness of the main results in Table~\ref{tab:main_results} to the exclusion of application outcomes recorded after the second QR code of the Requirements Letter broke due to an error, see fn.~\ref{footnote:error}. \\ See Table~\ref{tab:main_results} for more information. \\ 
      Standard errors are clustered at building address level. 
      \sym{*} \(p<0.05\), \sym{**} \(p<0.01\), \sym{***} \(p<0.001\)
      }
\end{table}

\revisedcolor 

\begin{table}[H]
    \scriptsize\singlespacing\centering
    \caption{\revisedcolor Comparison of marginal effects from OLS and logistic regression}\label{tab:logit_robustness}
    \medskip
        \def\sym#1{\ifmmode^{#1}\else\(^{#1}\)\fi}
        \begin{tabular}{l*{8}{c}}
        \hline\hline
            &\multicolumn{1}{c}{(1)}&\multicolumn{1}{c}{(2)}&\multicolumn{1}{c}{(3)}&\multicolumn{1}{c}{(4)}&\multicolumn{1}{c}{(5)}&\multicolumn{1}{c}{(6)}   \\
            &\multicolumn{6}{c}{\it Dependent variable: Naturalization application}\\
        \hline
        \multicolumn{6}{l}{\it Panel A. Coefficient estimates} \\
        \partialinput{7}{20}{revision/main_output_logit.tex} \\
        \hline 

        \hline
        \multicolumn{6}{l}{\it Panel B. Comparison of Policy tree with:}\\
        \partialinput{22}{29}{revision/main_output_logit.tex}
        \hline
        Sample & Wave 2 & Wave 1-2 & Wave 1-2 & Wave 2 & Wave 1-2 & Wave 1-2 \\
        \partialinput{30}{32}{revision/main_output_logit.tex}
        \hline\hline
       \end{tabular}
      \par\medskip
      \parbox{\linewidth}{
      \textit{Notes:} This table compares OLS and logistic regression estimation results.
      Columns (1)-(3) of this table reproduce columns (1)-(3) of the main results in Table~3 and rely on OLS. Columns~(4)-(6) are based on the same specifications but employ logistic regression and report the average marginal effects. Column~1 and 4 only use data from wave 2; the remaining columns use the full data set. The reference group in columns~1 and 4 is random treatment allocation; no treatment in the remaining columns. Panel A reports the coefficient estimates. Panel B compares the policy rule using policy trees against always assigning the same treatment to everyone and random treatment allocation. Covariates include the region of nationality, age, gender, years lived in Zurich and years lived in Switzerland. \\ 
      Standard errors are clustered at building address level. 
      \sym{*} \(p<0.05\), \sym{**} \(p<0.01\), \sym{***} \(p<0.001\)
      }
\end{table}

\revisedcoloroff

\end{document}